\documentclass[a4paper, 11pt, accepted=2025-08-16]{quantumarticle}
\pdfoutput=1
\usepackage[utf8]{inputenc}
\usepackage[english]{babel}

\usepackage[T1]{fontenc}
\usepackage[numbers,sort&compress]{natbib}

\usepackage{blindtext}
\usepackage{tikz}
\usetikzlibrary{quantikz2}
\usepackage{algorithm}
\usepackage{algpseudocode}
\usepackage{amsmath}
\usepackage{amssymb}
\usepackage{hyperref}
\usepackage{dsfont}
\usepackage{float}
\usepackage{graphicx}
\usepackage{tabularx}
\usepackage{subcaption}
\usepackage{comment}
\usepackage{soul}
\usepackage{tikz}
\usepackage{lipsum}

\begin{document}

\title{Reduced Sampling Overhead for Probabilistic Error Cancellation by Pauli Error Propagation}

\author{Timon Scheiber}
\orcid{0009-0006-0077-5350}
\email{timon.florian.scheiber@igd.fraunhofer.de}
\author{Paul Haubenwallner}
\orcid{0009-0004-2250-1866}
\author{Matthias Heller}
\orcid{0000-0002-4774-5072}
\affiliation{Fraunhofer Institute for Computer Graphics Research IGD} 
\affiliation{Technical University of Darmstadt, 
Interactive Graphics Systems Group}

\maketitle

\begin{abstract}
Quantum error mitigation is regarded as a possible path to near-term quantum utility.
The methods under the quantum error mitigation umbrella term, such as 
probabilistic error cancellation (PEC), zero-noise extrapolation (ZNE) or Clifford data regression (CDR) are able to significantly 
reduce the error for the estimation of expectation values, although at an exponentially
scaling cost, i.e., in the sampling overhead.
In this work, we present a method to reduce the sampling overhead of PEC through Pauli error propagation combined with classical preprocessing. Our findings indicate that this method significantly reduces sampling overheads for Clifford circuits, leveraging the well-defined interaction between the Clifford group and Pauli noise.
 Additionally, we show that the method is applicable to non-Clifford circuits, though with more limited effectiveness, primarily constrained by the number of non-Clifford gates present in the circuit. We further provide examples of Clifford sub-circuits commonly encountered in relevant calculations, such as resource state generation in measurement-based quantum computing.
\end{abstract}

\section{Introduction}

Quantum computing is expected to outperform classical computing in specific use cases within the near future~\cite{daley2022practical, di2023quantum}.
However, most of the existing algorithms showing a rigorously proven superior scaling compared to classical algorithms lie beyond the reach of current noisy intermediate scale quantum (NISQ) computers and will probably become relevant only after fault-tolerance is achieved \cite{gidney2021factor, reiher2017elucidating}.
While recently tremendous progress in the realization of error corrected qubits has been made, both in terms of efficient encodings~\cite{bravyi2023high, xu2024constant, bluvstein2024logical}, and real hardware demonstrations~\cite{bluvstein2024logical, sivak2023real, google2023suppressing, acharya2024quantum}, current quantum hardware is still far from being fault-tolerant.
On the other side, evidence has been presented that the current generation of quantum hardware can access computational spaces, which might be out of reach even for advanced supercomputers~\cite{morvan2023phase, kim2023evidence, shinjo2024unveiling}.
Since these devices are still bound by noise, current NISQ-algorithms require aid by quantum error mitigation (QEM) schemes to be able to compete with classical solutions \cite{czarnik2021error,temme2017error, cai2023quantum}.

QEM methods mostly focus on quantum algorithms, that aim to estimate the expectation value $\langle A \rangle$ of some observable $A$, by reducing the noise induced bias at the cost of an increase in the variance of the estimate~\cite{cai2023quantum, quek2024exponentially, takagi2022fundamental}. 
One of the earlier presented methods is the so-called probabilistic error cancellation (PEC)~\cite{temme2017error}. 
PEC aims to construct an ideal, noiseless circuit operation $\mathcal{U}(\rho) = U \rho U^\dagger$ by expanding it into an (over-complete) basis of natively performable, noisy operations $\mathcal{O}_i$, which can be directly executed by the hardware.
This expansion can be achieved in two ways:
Either by \textit{compensation}, where each gate operation $\mathcal{U}_i$ of a circuit $\mathcal{U} = \mathcal{U}_n \cdots \mathcal{U}_0$ is directly replaced by the superimposed operation $\mathcal{U}_i = \sum_j \eta_j \mathcal{O}_j$ for some real coefficients $\eta_j$,
or alternatively by \textit{inversion}, where for each noisy operation $\tilde{\mathcal{U}}_i = \Lambda_i \circ \mathcal{U}_i$ the mathematical inverse of the noise channel $\Lambda^{-1}_i= \sum_j \eta_j' \mathcal{O}_j$ is implemented directly before or after $\Lambda_i$ to cancel the effect of the noise~\cite{endo2018practical}.
The implementation of this decomposition is performed probabilistically by sampling from a quasi-probability distribution defined by the linear combination with probabilities respecting the weights $\eta_j$.
This implementation performs the ideal operation on average, but it generally comes at the cost of an increase in the variance of the desired result described by the sampling overhead $\gamma$. 
The latter method has recently been demonstrated experimentally on a superconducting quantum chip~\cite{van2023probabilistic}. 
In the reference, a sparse Pauli-Lindblad noise model was derived to efficiently characterize and learn the noise of sparsely connected quantum devices and estimate the inverse noise channels $\Lambda_i^{-1}$ for different layers of noisy two-qubit gates.
While the method delivers excellent results in terms of retrieving nearly bias-free estimates, the exponentially scaling increase in $\gamma$ still limits the usefulness of the method to small circuits.

In this work, we alleviate this scaling issue by introducing a method to estimate and sample from a conjoint inverse noise channel (or fused noise channel) $\Lambda_\text{global}^{-1}$ in contrast to sampling from each inverse noise channel separately.
To estimate this conjoint inverse noise channel we utilize Pauli error propagation through Clifford circuits. 
We show, that the proposed method can greatly reduce the required sampling overhead.

\subsection*{Related work and our main contributions}

Probabilistic error cancellation, often denoted quasi-probability decomposition method, is a widely utilized technique for error mitigation. The primary challenge in its applicability lies in the sampling overhead.
Numerous protocols have been developed to alleviate this issue \cite{tran2023locality, shyamsundar2025cv4quantum, zhao2024retrieving, jiang2021physical, guo2022quantum, filippov2023scalable, rennela2024low}.
For instance, Ref.~\cite{tran2023locality} demonstrates that for highly local observables, the sampling overhead can be reduced by leveraging the fact that information and, hence, noise propagation, is constrained within a light cone originating from the observable. 
Other work~\cite{shyamsundar2025cv4quantum} shows that the sampling overhead can be reduced by using the control variates method, commonly used in statistics for variance reduction of Monte-Carlo methods.
Ref.~\cite{zhao2024retrieving} introduces a novel method with reduced sampling overhead for extracting non-linear features, e.g. the $k$-th moment of a density operator $\text{Tr}(\rho^k)$, by measuring a shifted observable on $k$ copies of the original system.

In this work, we develop a method to reduce the sampling overhead for probabilistic error cancellation by leveraging the interference of local noise channels.
The interference from sequential error channels can result in decreased sampling overheads~\cite{jiang2021physical}, a principle that has been implicitly employed in recent tensor network based approaches~\cite{guo2022quantum, filippov2023scalable}.
Although tensor networks are not constrained by circuit width, they require approximations to capture deep circuits leading to a bias in the mitigated expectation values. 
In contrast, our method only cancels those terms, which interfere destructively without any approximations such that no bias is introduced.

We calculate the interference based on Heisenberg propagation of noise channels within the context of Clifford circuits and Pauli rotation gates.
By focusing our investigation on Clifford circuits and Pauli noise, we can effectively retrieve correlations from multiple layers of the quantum circuit.
Given that the reduction in sampling overhead arises from the build-up of interference over various layers, our method achieves significantly greater reductions than those possible with tensor networks which are usually applied to low depth circuits.
A further benefit of the Pauli error propagation method is that the inverse of the channels can be implemented by sampling from Pauli gates only and does not require any entangling gates, which might introduce more errors to the system.

A related work~\cite{rennela2024low} employs a similar approach of error propagation for the case of phase errors to reduce the sampling overhead of PEC.
However, in contrast to our approach, the authors consider cat qubits with dominant phase errors and a set of bias preserving gates and therefore only discusses the propagation of Pauli $Z$ errors.

\section{Probabilistic Error Cancellation for Pauli Errors}\label{Sec:PEC}
In this section we give a brief overview of PEC by inversion.
We assume that the individual noise channels occurring during a gate operation can be described as an ideal Pauli channel, which can be assured by randomized compiling over the Pauli group (also known as Pauli twirling)~\cite{wallman2016noise}
\begin{equation}
    \label{eq:Pauli_Channel}
        \Lambda(\rho) = \sum_{i=1}^{N = 4^n} c_{i} P_i \rho P_i^\dagger.
    \end{equation}
The summation runs over all elements $P_i$ of the Pauli group of dimension $n$, where $n$ denotes the number of qubits. 
The real, positive channel coefficients $c_i$ sum up to one and can be interpreted as the probability of a Pauli error $P_i$ occurring additionally to the effect of the ideal operation $\mathcal{U}$.
To perform PEC we assume that the exact coefficients of the correlated Pauli errors can be efficiently learned, for example by using techniques such as cycle benchmarking~\cite{van2023probabilistic, erhard2019characterizing}.

An important property of Pauli noise channels is that their mathematical inverse also closely resembles a Pauli channel 
\begin{equation}
\label{eq:inv_Pauli_Channel}
    \Lambda^{-1}(\rho) = \sum_{i=1}^N \tilde c_{i} P_i \rho P_i^\dagger,
\end{equation}
however, with coefficients $\tilde c_{i}$ that are no longer guaranteed to be positive.
While not being a channel in the mathematical sense, we will refer to the map defined by Eq.~\ref{eq:inv_Pauli_Channel} as the inverse channel throughout this manuscript.
The inverse is in general not a completely positive trace preserving (CPTP) map and can thus not be implemented by a single unitary operation.
Its effect can however be realized \textit{probabilistically}.

To implement the operation probabilistically, Eq.~\ref{eq:inv_Pauli_Channel} can be restructured as follows:
\begin{equation}
\label{eq:prob_implement}
    \Lambda^{-1}(\rho) = \gamma \sum_{i=1}^N \text{sgn}(\tilde c_{i}) \cdot p_i P_i \rho P_i^\dagger, \quad p_i = \frac{|\tilde c_{ i}|}{\gamma},
\end{equation}
where the factor $\gamma$ is given by
\begin{equation}
    \gamma = \sum_{i=1}^N | \tilde c_{i}|.
\end{equation}
The benefit of restructuring the inverse channel in this way stems from the fact that the coefficients $p_i$ can now be interpreted as probabilities.
Utilizing Eq.~\ref{eq:prob_implement}, the inverse is performed \textit{on average} by sampling and applying the Pauli correction $P_i$ directly before the noisy gate with corresponding probability $p_i$ and multiplying the measured expectation value with the corresponding sign ${\rm sgn}(\tilde c_i) = s_i$ and $\gamma$-factor in post-processing.

To apply the method to circuits containing multiple noisy gate operations (or layers of parallel executed operations) $\tilde{\mathcal{U}} = \prod_{l=1}^L (\mathcal{U}_l \circ \Lambda_l)$, this process is repeated for each noisy layer $\tilde{\mathcal{U}}_l$ of the circuit, so that an individual correction is drawn and directly applied in front of the layer.
Since each correction is attached to a corresponding sign and sampling overhead, the overall sign $s_\text{global} = \prod_l s_l$ and $\gamma$-factor $\gamma_\text{total} = \prod_l \gamma_l$ for the complete circuit are given by the product of the individual signs and sampling overheads respectively.

To retrieve the expectation value of the mitigated observable, several correction circuit instances are generated and executed.
The individual results are multiplied by the total sampling overhead $\gamma_\text{total}$, and the respective global sign $s_\text{global}$.
Finally, the mitigated expectation value is given by the average of the $M$ executed correction circuits~\cite{temme2017error}
\begin{equation}
    \langle A \rangle_\text{ideal} = \frac{\gamma_\text{total}}{M} \sum_{m=1}^M s_\text{global,m} \langle A \rangle_\text{corr,m}
\end{equation}
where $\langle A \rangle_\text{corr,m} = \text{Tr}(A~\tilde{\mathcal{U}}_\text{corr,m}(\rho))$ refers to the expectation value of $A$ evaluated on the $m$-th circuit.

As described by van den Berg et al.~\cite{van2023probabilistic}, the factor $\gamma \geq 1$ is directly related to the variance of the mitigated observable, which scales with $\gamma^2$. 
Since each noisy layer is associated with an individual noise channel, and thus an individual factor $\gamma_l$, the total $\gamma$-factor of the system increases exponentially. 
It follows that the variance of the mitigated expectation value and hence the number of shots required for a small sampling error, grows exponentially with the circuit depth.

\section{Propagated Probabilistic Error Cancellation (pPEC) for Clifford circuits}\label{Sec:pPEC}
One of the limiting factors of PEC is the large overhead to compensate for the rapid growth in variance.
We approach this problem by introducing a method to decrease the sampling overhead of classical PEC, which we call (error)-propagated probabilistic error cancellation (pPEC).
In pPEC the individual inverse noise channels of a given circuit are propagated to the start (or end) of the circuit and multiplied together, yielding a conjoint operation with smaller $\gamma$-factor.
To estimate this fused inverse noise channel, we present two methods; a hardware-agnostic Monte-Carlo method which serves as a more educational example (see appendix \ref{Appendix:MonteCarlo}) as well as an analytic method based on a sparse Pauli-Lindblad noise model \cite{van2023probabilistic}.
We motivate this approach by showing that it is favorable to sample from a conjoint inverse noise channel $\Lambda_{\text{global}}^{-1}$ by comparing the sampling overhead $\gamma_{\text{global}} = \gamma(\Lambda^{-1}_{\text{global}})$ of applying the global inverse (fused noise channel for all layers) to the sampling overhead of correcting each layer individually $\gamma_\text{total} = \prod_{l}  \gamma(\Lambda_l^{-1})$.

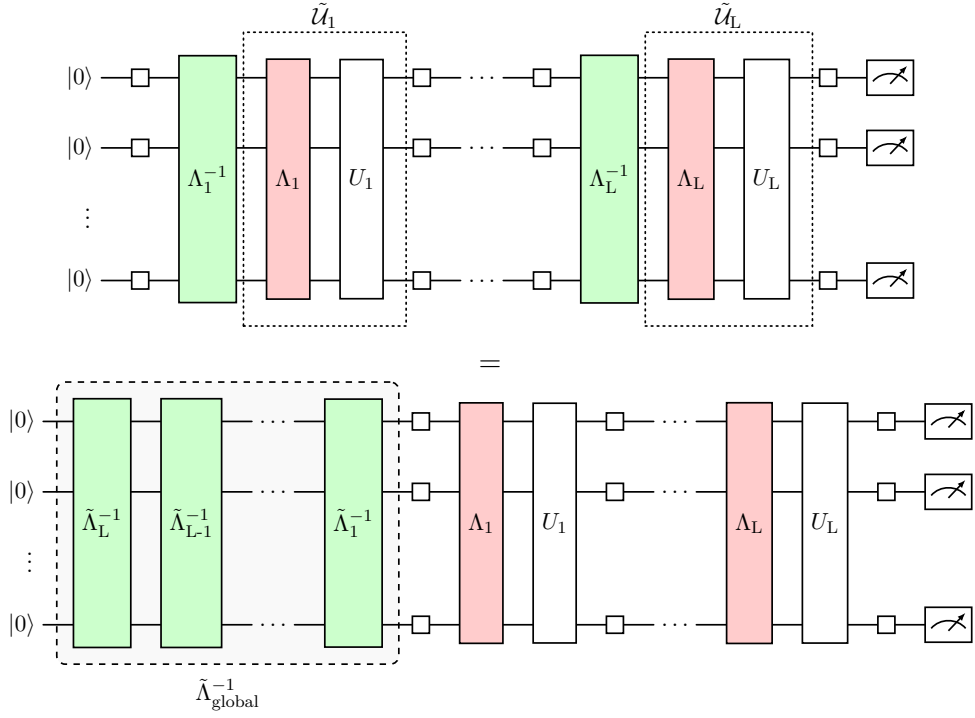
\begin{figure*}
\begin{center}
    \begin{tikzpicture}
    \node[scale=0.8] {
    \begin{quantikz}
        \ket{0} \ &\gate{}& \gate[4,style={fill=green!20}]{\Lambda_1^{-1}} &\gate[4, style={fill=red!20}]{\Lambda_{\text{1}}}\gategroup[4,steps=2, style={dotted,cap=round,inner sep=7pt}]{$\tilde{\mathcal{U}}_1$} &\gate[4]{U_{\text{1}}}  &\gate{}&  \ \ldots\ &\gate{}& \gate[4,style={fill=green!20}]{\Lambda_{\text{L}}^{-1}} &\gate[4, style={fill=red!20}]{\Lambda_{\text{L}}}\gategroup[4,steps=2, style={dotted,cap=round,inner sep=7pt}]{$\tilde{\mathcal{U}}_\text{L}$}  &\gate[4]{U_{\text{L}}}& \gate{}& \meter{}   \\
        \ket{0} \ &\gate{}&                                              &                                       &             &\gate{}&  \ \ldots\ &\gate{}&&&&\gate{}& \meter{}\\
        \ \vdots  &    \wireoverride{n}  &         \wireoverride{n}           &    \wireoverride{n}               &   \wireoverride{n}  &\wireoverride{n} & \wireoverride{n} &\wireoverride{n}&\wireoverride{n}&\wireoverride{n}&   \wireoverride{n}    & \wireoverride{n}\\
        \ket{0} \ &\gate{}&                                              &                                       &             &\gate{}& \ \ldots\  &\gate{}&&&&\gate{}& \meter{}
    \end{quantikz}
    };
    \end{tikzpicture}
    \\
        =
    \\
    \begin{tikzpicture}
    \node[scale=0.8] {
    \begin{quantikz}
        \ket{0} \ & \gate[4,style={fill=green!20}]{\tilde{\Lambda}_{\text{L}}^{-1}} \gategroup[4,steps = 4, style={dashed,rounded corners,fill=gray!4.5}, label style = {label position=below, yshift=-0.7cm}, background]{$\tilde{\Lambda}_{\text{global}}^{-1}$} &\gate[4, style={fill=green!20}]{\tilde{\Lambda}_{\text{L-1}}^{-1}} & \ \ldots\ &\gate[4, style={fill=green!20}]{\tilde{\Lambda}_{\text{1}}^{-1}}  &\gate{}  &\gate[4, style={fill=red!20}]{\Lambda_{\text{1}}} &\gate[4]{U_{\text{1}}}& \gate{}& \ \ldots\ &\gate[4, style={fill=red!20}]{\Lambda_{\text{L}}} &\gate[4]{U_{\text{L}}}& \gate{}& \meter{}   \\
        \ket{0} \ &                                              &                                                      & \ \ldots\ &                                                 &\gate{}  &                                         &             &\gate{}& \ \ldots\  &                                         &             & \gate{}&  \meter{}\\
        \ \vdots  &    \wireoverride{n}  &         \wireoverride{n}           &    \wireoverride{n}               &   \wireoverride{n}  &\wireoverride{n} & \wireoverride{n} &\wireoverride{n}&\wireoverride{n}&\wireoverride{n}&   \wireoverride{n}    & \wireoverride{n}\\
        \ket{0} \ &                                              &                                                      & \ \ldots\ &                                                 &\gate{}  &                                         &             &\gate{}& \ \ldots\  &                                         &             & \gate{}&  \meter{}
    \end{quantikz} 
    };
    \end{tikzpicture}
\end{center}
\caption{\textbf{Above:} graphical summary of the PEC method. The circuit consists of one qubit gates and
layers of noisy two qubit gates $\tilde{\mathcal{U}}_l = \mathcal{U}_l \circ \Lambda_l$. The noise is probabilistically canceled 
by implementing the inverse of the noise channel and applying it in front of the noise.
\textbf{Below:} the same circuit after propagation of the inverse noise channels to the start of the circuit.
After multiplication the full noise of the circuit is canceled with the single layer correction $\Lambda_{\text{total}}^{-1}$.}\label{pPEC:circuit}
\end{figure*}

In a naïve approach to estimate the global noise channel, one could try to directly apply a learning procedure to the full circuit which, in the general case, would require an exponentially scaling amount of measurements and is thus infeasible.
However, the circuit can be decomposed into a product of individual noisy gate layers $\tilde{\mathcal{U}}_l$, each consisting of a (largely) learnable~\cite{chen2023learnability} noise channel $\Lambda_l$. 
It is customary to assume that single-qubit operations are effectively noiseless and only two-qubit operations contribute to the noise due to order of magnitude higher errors on most devices.
The full circuit operation $\tilde{\mathcal{U}}$ can then be decomposed into a product of noisy circuit operations $\tilde{\mathcal{U}}_l$
\begin{equation}
    \tilde{\mathcal{U}} = \prod\limits_{l} \tilde{\mathcal{U}}_{l} = \prod\limits_{l} (\mathcal{U}_l \circ \Lambda_l),
\label{eq:circuit_decomposition}
\end{equation}
where we assume $\Lambda_l = \mathds{1}$, if $\mathcal{U}_l$ consists solely of single-qubit gates.
The effect of the noise can be canceled by probabilistically implementing the inverse of the noise channel $\Lambda_l^{-1}$ directly before the application of the noisy gate $\tilde{\mathcal{U}}_l$. 
A graphical depiction of the method is presented in the upper part of Fig.~\ref{pPEC:circuit}.

Since we are only interested in calculating the global inverse noise channel and not the global noise channel itself, we consider the PEC-corrected circuit $\mathcal{U}(\rho) = \prod_{l=1}^L \mathcal{U}_l  \Lambda_l \Lambda_l^{-1}$ as the point of departure.
From this we can calculate the global inverse by propagating each inverse to the start of the circuit
\begin{equation}
    \Lambda^{-1}_\text{global} = \prod_{l=L}^{1} \tilde{\Lambda}_l^{-1},
\end{equation}
where the individual, propagated operations are given by
\begin{equation}
    \tilde{\Lambda}_l^{-1} = \mathcal{U}_0 \cdot \cdot \cdot \mathcal{U}_{l-2}\mathcal{U}_{l-1}(\Lambda_l^{-1}),
\end{equation}
with $\mathcal{U}_j(\Lambda_l^{-1}) =  U_j \Lambda_l^{-1} U_j^\dagger$ describing the Heisenberg propagation of the inverse channels.
Note that the product runs over the layers in reversed order, and we only conjugate the inverse noise channels with the ideal circuit operations and not the noise channels.

It is important to note, that for arbitrary gates the inverse after the propagation might no longer be described by a diagonal (inverse) Pauli channel.
An important class of gates, for which the inverse noise channel will remain a diagonal Pauli channel are Clifford gates.
This is the case, since elements of the Clifford group are defined by the property 
\begin{equation}
\label{eq:CliffordCommute}
    CPC^\dagger = P',
\end{equation}
where $P$ and $P'$ denote operators of the Pauli group and $C$ an arbitrary operator of the Clifford group.
Using Eq.~\eqref{eq:CliffordCommute} we are able to move the Pauli correction terms in front of a Clifford gate by exchanging it with the conjugated Pauli correction $P'$
\begin{equation}
    \begin{quantikz}
        \qw &\gate[2]{C} & \gate{P_1} &\\
        \qw &            & \gate{P_2} &
    \end{quantikz}
    =
    \begin{quantikz}
        \qw & \gate{P_1'} & \gate[2]{C} &\\
        \qw & \gate{P_2'} &             &
    \end{quantikz}.
\end{equation}
In general this conjugation can be performed efficiently by using a look-up table, which we provide in Appendix~\ref{App:Tables} for the convenience of the reader. 
This way the individual channels can be propagated to the start of the circuit and multiplied together yielding the global (inverse) noise channel.

The key advantage of fusing individual correction layers together into one global one stems from the fact, that different errors or corrections can interfere destructively.
The identification of these corrections before applying the inverse probabilistically can thus reduce the required number of correction circuits by a sometimes large amount.
For example, the PEC corrected expectation value with $M$ independently sampled corrections can be expressed as
\begin{equation}
\label{eq:PEC_expval}
 \langle A \rangle_{\text{PEC}} = \frac{\gamma_\text{total}}{M} \left[ \sum_{m=1}^M s_m \langle A \rangle_{\text{corr,m}}  \right],
\end{equation}
where $\langle A \rangle_{\text{corr,m}}$ denotes the expectation value of $A$ on the $m$-th circuit and $s_m$ the corresponding global sign.
By identifying the number of corrections $\mathcal{J}$ that interfere destructively, i.e., those for which the local corrections map to the same global correction with opposite sign, Eq.~\eqref{eq:PEC_expval} can be restructured as
\begin{align}   
 \langle A \rangle_{\text{PEC}} &= \frac{\gamma_\text{total}}{M} \left[ \sum_{m=1}^{M-\mathcal{J}} s_m \langle A \rangle_{\text{corr,m}}  + \underbrace{\sum_{j=1}^\mathcal{J} s_j \langle A \rangle_{\text{corr,j}}}_{0} \right]\nonumber\\
                                 &= \frac{\gamma_\text{total}}{M} \left[ \sum_{m=1}^{M-\mathcal{J}} s_m \langle A \rangle_{\text{corr,m}} \right],
\label{eq:reduced_exp_val}                    
\end{align}
where, by assumption, the last $\mathcal{J}$ correction circuits interfere destructively meaning that the individual expectation values cancel each other out.
In contrast, by only sampling from the non-interfering corrections, the pPEC corrected expectation value with $M-\mathcal{J}$ independently drawn correction circuits is given by 
\begin{equation}
    \langle A \rangle_{\text{pPEC}} = \frac{\gamma_\text{total}}{M-\mathcal{J}} \left[ \sum_{m=1}^{M-\mathcal{J}} s_m \langle A \rangle_{\text{corr,m}} \right],
\end{equation}
which coincides with Eq.~\eqref{eq:reduced_exp_val} up to a factor of $\frac{M-\mathcal{J}}{M}$.
This factor can then be incorporated into the $\gamma$-factor giving the sampling overhead for pPEC
\begin{equation}
    \label{eq:gamma_ppec}
    \gamma_{\text{pPEC}} = \gamma_{\text{total}} \cdot \frac{M-\mathcal{J}}{M}.
\end{equation}

Hence, sampling from the pPEC distribution can be interpreted as sampling from a quasi-probability distribution with reduced overhead $\gamma$.
The reduction stems from the fact that pPEC is able to identify the interfering corrections before the implementation.
We provide a more illustrative example of how the interference of corrections follows from error propagation as well as a method to estimate a fused channel in Appendix~\ref{Appendix:MonteCarlo}. 
Furthermore, we provide a more rigorous proof that the sampling from the global or fused inverse is always favorable in Appendix~\ref{Appendix:Proof}.

\subsection{XI-Reduction}
\label{sec:XI-red}
To decrease the sampling overhead even further, we reduce the number of individual Pauli corrections that need to be applied at the start of the circuit, by leveraging the phase invariance of the computational basis states.
Considering for example the Pauli $Z$ operator it is easy to verify that $Z  \ket{0} = \ket{0}$ and $Z \ket{1} = -\ket{1} \equiv \ket{1}$, up to a global phase.

This allows us to replace each $Z$ operator with an identity operation if applied to a computational basis state without superposition, as for example at the start of the circuit. 
Furthermore, the Pauli $Y$ operator can be expressed as $Y =-i X \cdot Z$ and thus $Y  \ket{0} \equiv X \ket{0}$ and $Y \ket{1} \equiv X \ket{1}$, again omitting the global phase.
These identities also hold for tensorized Paulis. For example, the Pauli strings $IXXZY \equiv IXXIX$ and $IYYIX \equiv IXXIX$ are both equivalent to the same Pauli string after the reduction and can thus be corrected by the same operation.
Using these reductions the total number of corrections and thus the maximum number of instances that need to be sampled individually reduces from $4^n$ to $2^n$.

It is noteworthy that this reduction is inherently symmetric, in the sense that it can be applied either after propagation of the gates to the start or to the end of the circuit, directly before the measurement.
This convenience directly follows from Born's rule, stating that only the squared magnitude of the amplitudes can be measured.
For example in the computational basis the measurement of a quantum state $\ket{\psi} = \sum_{i \in \{0,1\}^n} c_i \ket{i}$ will result in the outcome $i$ with probability $|c_i|^2$.
Thus, any transformation $c_i \rightarrow e^{i\phi_i} c_i$ does not change the measurement outcome.
Note, that the final measurement is always in the computational basis, when incorporating the final basis change into the circuit.

\subsection{Readout-Error-Mitigation}
\label{subsec:REM}
Another common source of errors in quantum circuits are state preparation and measurement (SPAM) errors.
Here, we utilize PEC for the mitigation of measurement errors and integrate them into the pPEC workflow. 
The readout error mitigation method presented in this work is based on assignment matrix inversion methods~\cite{bravyi2021mitigating, nation2021scalable, funcke2022measurement}.
In these methods an assignment matrix $A$ is calculated by preparing each basis state individually and measuring the noisy outcome.
The matrix $A$ correlates the ideal measurements with the noisy measurements via the relation $\vec{p}_{\text{noisy}} = A \vec{p}_{\text{ideal}}$.
The idealized counts can then be retrieved by inverting the assignment matrix and multiplying the experimentally retrieved probability vector of the bit-strings with the inverse $\vec{p}_{\text{ideal}} = A^{-1} \vec{p}_{\text{noisy}}$.
While this method almost fully eliminates readout errors, an exponential overhead for the preparation of the $2^n$ basis states is introduced, rendering the method infeasible for larger circuits.

To avoid this overhead we assume a tensor product noise model as presented in Ref.~\cite{bravyi2021mitigating}
\begin{equation}
\label{eq:assignment}    
    A = \bigotimes_{i=1}^n A^i = \bigotimes_{i=1}^n \begin{bmatrix}
        1-\epsilon_i & \eta_i \\
        \epsilon_i & 1-\eta_i 
        \end{bmatrix},
\end{equation}
with $\epsilon_i$ and $\eta_i$ denoting the $\ket{0} \rightarrow 1$ and $\ket{1} \rightarrow 0$ faulty readout probabilities on qubit $i$ respectively.
This imposes a significant simplification, but it can be considered reasonable for sparse measurement outcomes~\cite{yang2022efficient}.
The tensor product of Eq.~\eqref{eq:assignment} runs over the individual $2 \times 2$ assignment matrices, which are defined for each individual qubit.
These matrices are generally not equal among the qubits nor symmetrical under transposition, meaning that the individual bit-flip probabilities for $\ket{0} \rightarrow 1$ and $\ket{1} \rightarrow 0$ are not identical.

To apply the assignment matrix method in a fashion that integrates to the pPEC framework, we need to construct a measurement error channel that closely resembles a Pauli channel. 
This is achieved by symmetrizing the matrices $A^i$ via randomized insertion of $X$ gates~\cite{van2022model}, i.e.,
an $X$ gate is applied with 50\% probability directly before the measurement for each qubit and corrected in post-processing.
This procedure symmetrizes the matrix $A^i_\text{twirl}$ = $(A^i_\text{twirl})^T$ (see appendix~\ref{appendix:readout}) which can then be incorporated into our framework using a Pauli $X$ channel:
\begin{equation}
    \Pi_{\text{meas}}(\rho) = \sum_m (1-p_x) \pi_m I \rho I  \pi_m + p_x \pi_m X \rho X \pi_m,
\end{equation}
where $\pi_m$ are projection operators, fulfilling the relations $\pi_m \pi_{m'} = \delta_{mm'}\pi_m$, $\pi_m = \pi_m^\dagger$ and $\sum_m \pi_m = I$.

The bit-flip probabilities are explicitly given by the arithmetic average of the individual bit-flip probabilities, $p_{x, i} = \frac{\epsilon_i + \eta_i}{2}$.

\subsection{Classical cost}
We now discuss the classical cost of calculating the global or fused inverse noise channel.
Our algorithm consists of two main steps, the channel propagation and the channel product.
We first consider the cost associated with the propagation.

The commutation of a single Pauli with a Clifford gate can be performed by a look-up table in $\mathcal{O}(1)$.
This propagation must be performed for all Paulis in a given channel, which can still be achieved with constant overhead, with the constant  depending on the number of Paulis present in the channel.
Since we only consider Pauli channels and Clifford gates, the total number of operators that need to be stored is conserved.
The cost of propagating a single layer to the start of the circuit can be estimated  as the overhead of commuting a noise channel of $n$ Paulis by $d$ layers.
Hence, the propagation overhead for all Paulis scales polynomially as $\mathcal{O}(n \cdot d^2)$ due to each layer's propagation to the circuit start with incrementally increasing depth $d$.

The bottleneck of our algorithm is the calculation of the product of $k$ Pauli channels, where the number of terms, that one takes into account, can be chosen dynamically to restrict the computational time. 
This results in a general trade-off between classical and quantum resources.
Taking the product of two Pauli channels $\sum_i^N c_i P_i \rho P_i$ with $N$ respective elements, takes $N^2$ separate multiplications.
The cost of fusing $k$ channels can thus be upper bounded by $\mathcal{O}(N^k)$ which results in an exponential scaling in the classical cost.
However, this scaling is a worst-case scaling without regarding interferences between channels.

For instance, the product of $k$ single qubit noise channels acting on the same qubit seems to result in a $\mathcal{O}(4^k)$ scaling.
Since the resulting operators of the multiplication are restricted to the $4$ possible Pauli terms, in each step the corrections can be expressed as the $4$ possible Pauli corrections. Hence, the scaling is given by taking $k \cdot 4^2$ multiplications, with overhead $\mathcal{O}(k)$ far fewer than the projected $4^k$ terms. This reduction in exponential overhead applies to any channel multiplication. The overhead is minimized when many corrections map to the same operator after an expansion step. As this effect is responsible for interference, multiplying channels with large shared support will lead to minimal classical overhead while reducing the quantum overhead significantly.

Up to this point the calculations presented here have been exact, preserving the bias-free property of PEC.
If one allows for at least some small bias, the classical cost can be drastically reduced.
The magnitude of the coefficients $|c_i|$ of the inverse channel $\Lambda^{-1}(\rho)$ can be regarded as the probability of applying a specific correction.
Due to the multiplication of several corrections these values will be highly attenuated for some corrections while other coefficients will build up.
By truncating all coefficients smaller than a specific cutoff $\epsilon$, the truncated inverse channel can be expressed as
\begin{equation}
    \Lambda^{-1}_\text{trunc}(\rho) =\sum_j \Theta(|c_j| - \epsilon)  c_j P_j \rho P_j 
\end{equation}
where $\Theta(c_j)$ denotes the Heaviside step function and $c_j$ is the coefficient of Pauli $j$.

Truncating the coefficients at each step introduces a slight bias but significantly reduces the overhead in the required multiplications so that smaller sampling overheads can be achieved.

\section{Numerical simulations}
Due to the large shot budget and time required to fully characterize the complete noise channels we restrict the experiments to numerical simulations.
Since pPEC does not rely on approximations it is expected that the performance will be similar to regular PEC.

\subsection{pPEC for gate-level noise}\label{sec:results}
To demonstrate the efficiency of our proposed method, we compare PEC and pPEC for a random 10 qubit circuit consisting of Hadamard ($H$), Pauli ($X$,$Y$,$Z$), phase ($S$), controlled phase ($C_Z$) and measurement operations under gate level noise.
For each two-qubit gate we apply a depolarizing channel
\begin{equation}
    \Lambda_{\text{depol}}(\rho) = (1 - p) I \rho I + \frac{p}{(N-1)} \sum_{i=2}^{N=4^2} P_i \rho P_i^\dagger,
\end{equation}
where we chose the error probability $p = 2\%$ to resemble current hardware error rates \cite{van2023probabilistic}.
It should be noted that processes such as hardware specific idle times of qubits which generally lead to dephasing and decay are not modeled in this description.
Additionally, we integrate a tensor product, asymmetric readout error with a mean of $p_\text{read} = 2\%$, which we symmetrize using the method we introduced in Sec.~\ref{subsec:REM}. 

Using a random Clifford circuit we benchmark the three different PEC protocols -- PEC, pPEC and pPEC with XI-reduction.
We perform the method by drawing corrections from each inverse noise channel directly (PEC) or by drawing from the pPEC distribution.
For the considered example the pPEC circuits have been calculated by the direct product of all channels, which is generally possible for circuits with modest numbers of qubits.
For larger circuits the same can be done, however the product needs to be truncated as the space of possible corrections grows exponentially. 
From each of these distributions a total of 40 correction circuits is drawn, which are simulated with $1024$ shots per circuit instance.

\begin{figure}
    \includegraphics[width=.5\textwidth]{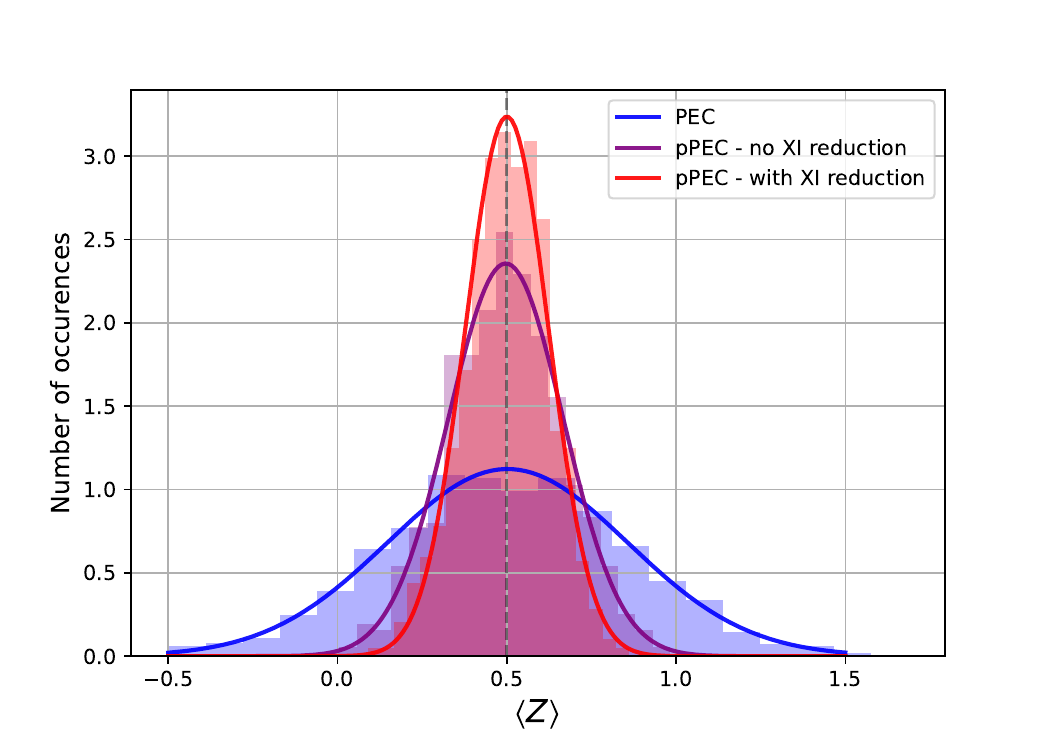}
\caption{Mitigated circuits for the three tested method, PEC (blue), pPEC (purple), pPEC with XI-reduction (red). All methods are able to retrieve error free estimates of the noiseless observable.
pPEC generally leads to smaller variances than the direct approach.}\label{figure1}
\end{figure}

To obtain a sufficient amount of statistics, a sample with a total of $1000$ mitigated expectation values per method was generated.
The resulting distribution is presented in Fig.~\ref{figure1}.
As expected, all PEC methods are able to retrieve the bias free expectation value on average, although at vastly different variances.
Due to the promising results we now explore the scaling of the different approaches.

\begin{figure}[h]
    \includegraphics[width=.5\textwidth]{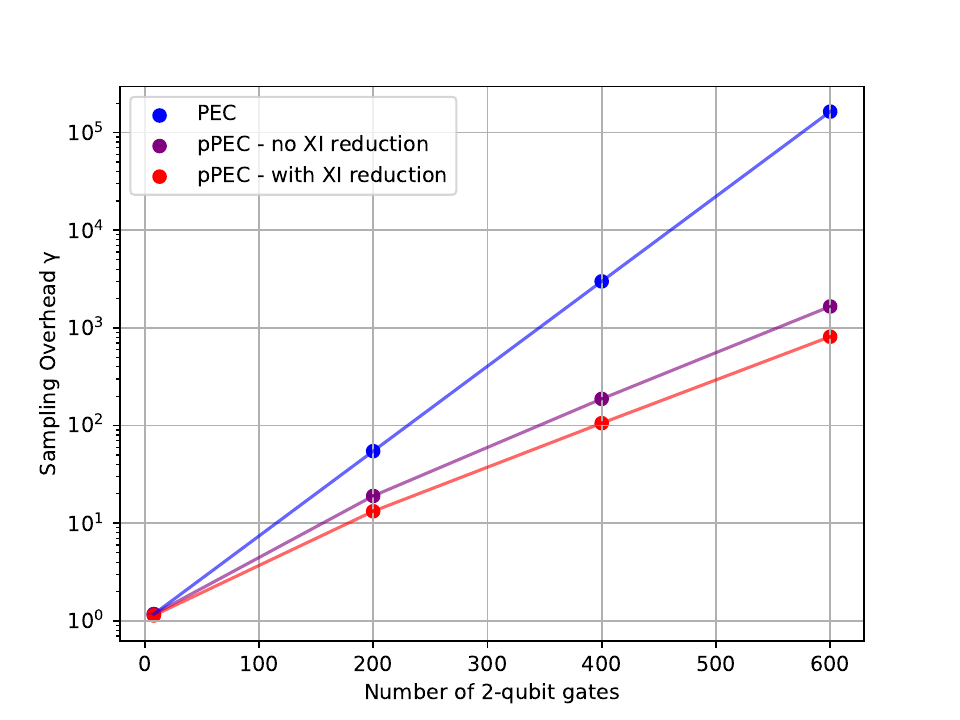}
\caption{Calculated Sampling overhead averaged over 100 instances of random 5-qubit Clifford circuits at an increasing number of noisy operations.
}
\label{figure2}
\end{figure}

To investigate the expected scaling, we benchmark the different methods by calculating the global inverse for a sample of 100 random 5-qubit Clifford circuits for an increasing number of noisy two-qubit operations and calculate the expected $\gamma$-factors.
The estimated total $\gamma$-factors for the direct PEC, propagated PEC and propagated PEC with XI-reduction are presented in Fig.~\ref{figure2}. 
Since the sampling-overhead scales exponentially in the number of noisy operations, the data is plotted logarithmically. 
We achieve a far more favorable scaling for the propagated approach and a further decrease with the XI-reduction.
However, it should be noted that the expected reduction can generally depend on the noise and structure of the circuit.
\subsection{pPEC for circuit-level noise with SPL models}
Sparse Pauli-Lindbladian (SPL) models are a recently introduced form of noise model to efficiently capture the noise of devices with limited physical qubit connectivity~\cite{van2023probabilistic}.
In the SPL model the noise is generated by a dissipative Lindbladian of the form 
\begin{equation}
\mathcal{L}(\rho) = \sum_{k \in \mathcal{K}} \lambda_k (P_k \rho P_k^\dagger - \rho),
\end{equation}
with model coefficients $\lambda_k$ and Lindblad jump operators $P_k$ given by a sparse subset $\mathcal{K}$ of the Pauli group.
A noise channel, defined for a full layer of in-parallel executable gates, is described by the formula
\begin{equation}
    \Lambda (\rho) = \prod_{k \in \mathcal{K}}(w_k \mathcal{I}(\cdot) + (1-w_k) \mathcal{P}_k(\cdot)) \rho,
\label{eq:SPL_noise}
\end{equation}
where the coefficients $w_k$ are given by $w_k = (1+e^{-2\lambda_k})/2$.
The product runs over the indexed set $\mathcal{K}$ with usually far fewer terms than the dense noise model $|\mathcal{K}| \ll 4^n$.

$\mathcal{K}$ can be chosen in a way that only Pauli strings with no identity terms on at most two (physically connected) qubits are present.
$\mathcal{P}_i(\cdot) \overset{\scriptscriptstyle\wedge}{=} P_i \cdot P^\dagger_i$ is a shorthand notation for a single Pauli channel in Kraus representation and the $(\cdot)$ symbol is a placeholder to illustrate that the whole product is applied to the system's density operator $\rho$.

In the following, we simulate a device with linear qubit topology, i.e., we choose the model Paulis in Eq.~\eqref{eq:SPL_noise} by choosing weight-two (and one) Pauli strings with non-identity terms only on nearest neighbor qubits.
As noise model we consider a quasi depolarizing channel by choosing the coefficients $w_k$ to be homogeneous, allowing for easy tuning of the noise strength.
We adjust the strength in accordance to the highest reported Pauli fidelity reported in Ref.~\cite{van2023probabilistic}, $f_\text{max} \approx 0.996$, by utilizing $w_k = \frac{1 + f_k}{2}$ where $f_k = \frac{1}{2^n}\text{Tr}(P_k^\dagger \Lambda(P_k))$ is the Pauli fidelity for Pauli $P_k$ with respect to the noise channel $\Lambda$.

\begin{figure}
    \includegraphics[width=.5\textwidth]{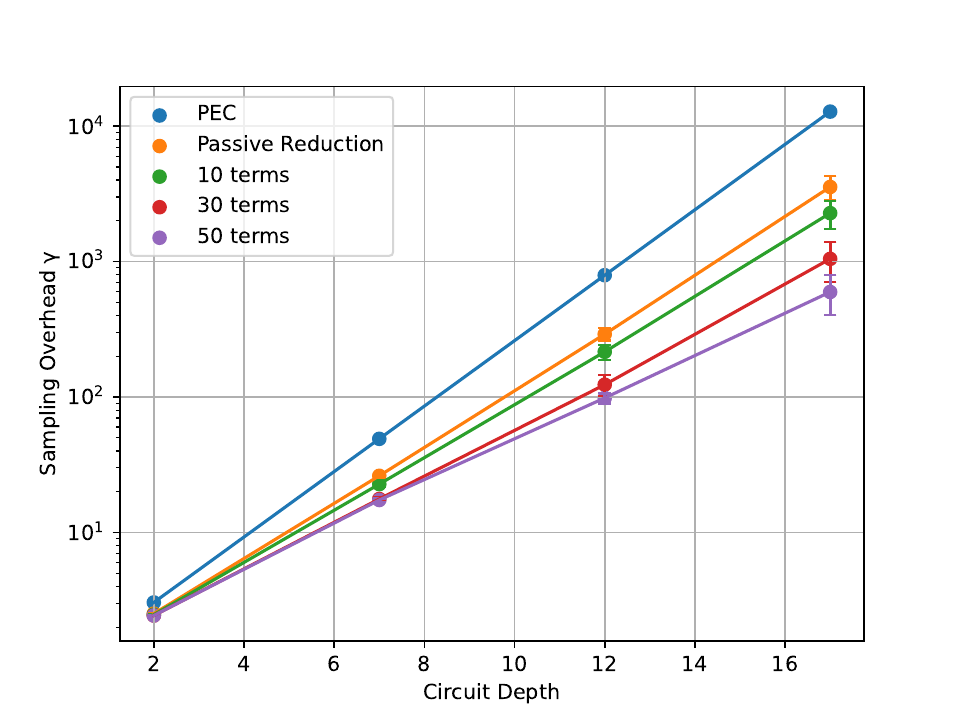}
\caption{Calculated $\gamma$ factors for quantum circuits of increasing circuit depth consisting of $10$ qubits with an average Pauli fidelity of $f_\text{avg} = 0.996$.
The results are averaged over 100 instances of random circuit instances, error bars indicate one standard deviation. 
The more channels are expanded, the higher the reduction in the sampling overhead $\gamma$ is achieved, however at an increase in the classical preprocessing cost.}\label{fig:SPL_pPEC}
\end{figure}

\begin{figure*}
    \begin{subfigure}[h]{0.49\linewidth}
    \includegraphics[width=\linewidth]{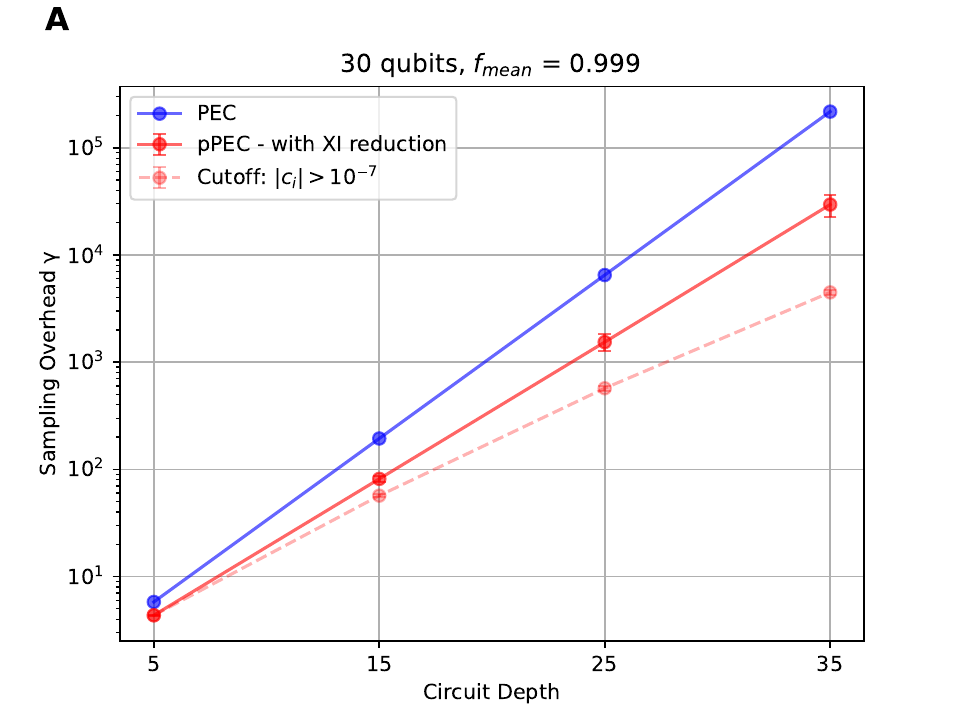}
    \end{subfigure}
    \hfill
    \begin{subfigure}[h]{0.49\linewidth}
    \includegraphics[width=\linewidth]{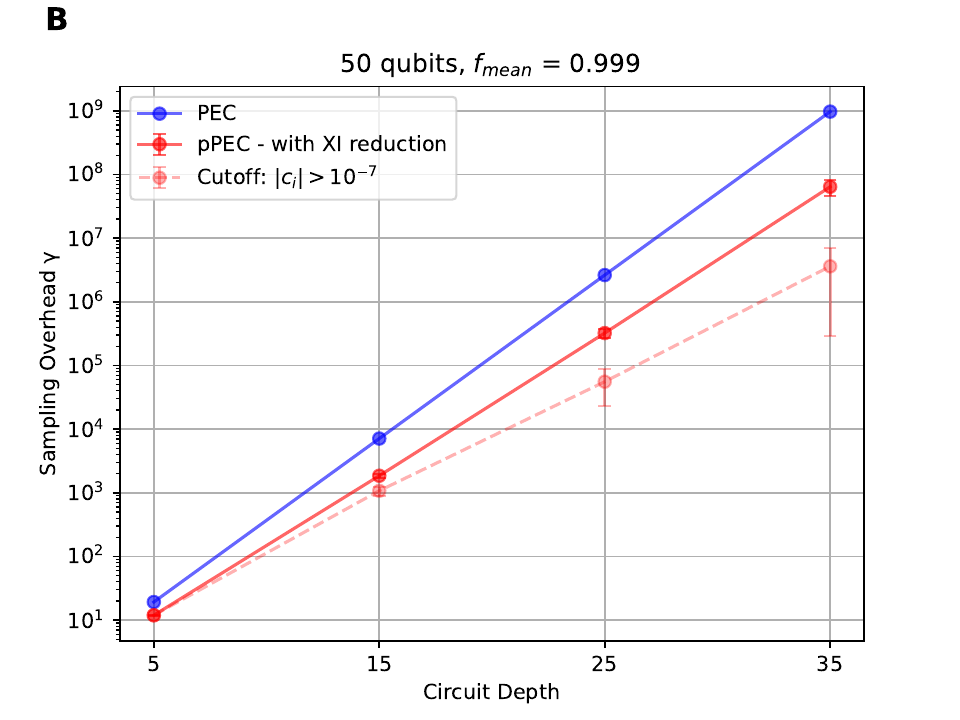}
    \end{subfigure}%
    \caption{Performance of pPEC with XI reduction for future devices at 30 qubits (\textbf{A}) and 50 qubits (\textbf{B}).
    The results are averaged over 100 random circuit instances with error bars indicating one standard deviation.
    The amount of expanded channels was taken dynamically, ranging from 0 (lowest depth) to 200 (highest depth). The dashed curves show approximate versions of pPEC where channel coefficients with a magnitude smaller the $10^{-7}$ have been cut off. Since the reduction greatly reduces the number of channels, the number of expansion steps was increased by a factor of 4 for all calculations. While this method is more computationally efficient it generally comes at the cost of a bias.}
    \label{fig:SPl_pPEC_higher_qubits}
\end{figure*}

We present our results of pPEC with XI reduction for random Clifford circuits of increasing depth in Fig.~\ref{fig:SPL_pPEC}. 
In contrast to the direct multiplication of individual channels, we first construct the global inverse channel in product form, which we expand term by term to reduce the sampling overhead.
We find that for the SPL model the XI reduction leads to the vanishing of terms which contain only $Z$ and $I$ operations, which we denote as \textit{passive reduction}.
The details on how pPEC is applied to the SPL noise model are presented in Appendix~\ref{Appendix:SPL}.

It is important to note that the ordering in which this expansion is performed is crucial to obtain meaningful reductions.
Due to the large sampling overhead for the considered device noise we limit the investigation to $10$ qubits.
The data for higher qubit counts and lower error rates – as expected for future generations of hardware – is presented in Fig.~\ref{fig:SPl_pPEC_higher_qubits}.

Fig.~\ref{fig:SPL_pPEC} shows a clearly favorable scaling relative to direct PEC, reaching an improvement of an order of magnitude even for modest depths.
On the other hand a direct comparison to Fig.~\ref{fig:SPl_pPEC_higher_qubits} shows that the effect of pPEC is far less drastic for lower error rates.
We attribute this to the fact, that the amount of interference is generally proportional to the magnitude of the channel coefficients $(1-w_k)$.
For lower error rates the magnitude will not be as large and thus deeper circuits are required to achieve comparable results.
However, even for larger circuits and modest depths a noticeable reduction in the sampling overhead can be achieved. 
We therefore conjecture that pPEC is a practical approach to reduce the amount of quantum resources, as it is always favorable in comparison to direct PEC.\\

We now consider possible relevant applications in which pPEC might be applicable and help reduce the workload of the quantum computer.

\section{Applications}

As described in the prior sections, the presented method is best suited for quantum circuits consisting of Clifford gates.
While Clifford circuits themselves are not of any practical relevance due to their efficient simulability~\cite{gottesman1998heisenberg}, larger Clifford structures do occur in many quantum circuits.
In this section we give a brief overview of some possible applications of the proposed method.

\subsection{Clifford structures in quantum circuits}

\subsubsection{Clifford building blocks}
Clifford sub-circuits frequently occur in quantum circuits. 
In this section, we focus on two fundamental building blocks that often arise in quantum algorithms; the $C_X$-ladder as well as layers composed of SWAP gates. 
The $C_X$-ladder is commonly employed to implement multi-qubit rotations, which are integral to many circuits, as for example the Quantum Approximate Optimization Algorithm (QAOA) \cite{farhi2014quantum}.
The general structure is given as
\begin{equation}
    \begin{quantikz}
        \qw   & \ctrl{1} &           &          &         &\\
        \qw   & \targ{}  & \ctrl{1}  &          &         &\\
        \wave &          &           & \ctrl{1}         &         &\\
        \qw   &          &           &\targ{1} & \ctrl{1}&\\
        \qw   &          &           &          & \targ{} &
    \end{quantikz},
\end{equation}
which consists of several $C_X$ gates applied in series.
Due to the sequential structure, the gates cannot be easily applied in parallel, rendering them ideal for a pPEC based approach.

Another often occurring Clifford building block is given by SWAP layers.
For devices with sparse qubit connectivities, as for example superconducting quantum computers, interactions of physical qubits that are spatially separated need to be bridged by SWAP gates.
SWAP gates are elements of the Clifford group as they can be decomposed to three consecutive $C_X$-gates:
\begin{equation}
\begin{quantikz}
    \qw & \swap{1} & \qw \\
    \qw & \targX{} & \qw
\end{quantikz}
=
\begin{quantikz}
    \qw & \ctrl{1} &\targ{} & \ctrl{1} & \qw \\
    \qw & \targ{}  & \ctrl{-1} & \targ{}  & \qw
\end{quantikz}.
\end{equation}

Since these gates cannot be applied in parallel each $C_X$-gate would contribute with an individual noise channel $\Lambda_i$, which can be reduced with pPEC.
Furthermore, SWAP gates are favorable for pPEC because the conjugation of a model Pauli does not change the weight of the Pauli string, i.e., the amount of non-identity terms in a Pauli string stays invariant (see Appendix~\ref{App:Tables}).

\subsubsection{Reducing state preparations in VQE}
A further interesting possible application for near-term quantum algorithms is the minimization of the number of state preparations in the Variational Quantum Eigensolver (VQE) \cite{peruzzo2014variational}.
The VQE algorithm aims to estimate the minimum eigenvalue of a given hermitian matrix $H$ by preparing a parametrized state $\ket{\psi(\theta_1, \theta_2, ..., \theta_n)} = \ket{\psi(\boldsymbol{\theta})}$ and iteratively minimizing the measured energy
\begin{equation}
    E = \underset{\boldsymbol{\theta}}{\text{min}} \bra{\psi(\boldsymbol{\theta})} H \ket{\psi(\boldsymbol{\theta})}
\end{equation}
by optimizing the parameters of the trial wave function.
The optimization is performed via a classical optimization loop, whereas the quantum computer is used to estimate the energy of each generation of parameters.

A limitation of this method is the required number of state preparations needed to measure a single instance of $\bra{\psi(\boldsymbol{\theta})} H \ket{\psi(\boldsymbol{\theta})}$.
Usually the matrix $H$ is expressed as a weighted Pauli sum $\sum_j c_j P_j$ so that the total energy estimate reduces to estimating each individual Pauli expectation value separately.
For a large number of Pauli terms the number of required measurements, and therefore state preparations, can rapidly become prohibitive.

To circumvent this problem, Ref.~\cite{gokhale2019minimizing} introduces a method which aims to reduce the number of state preparations by grouping the Pauli operators into commuting groups, which can be measured in parallel.
To measure the operators using a single state preparation,a basis change is performed, which is appended to the state-preparation circuit.
The basis change circuit consists entirely of Clifford gates, as is expected for a mapping from Pauli operators to Pauli operators rendering these circuits ideal for pPEC.
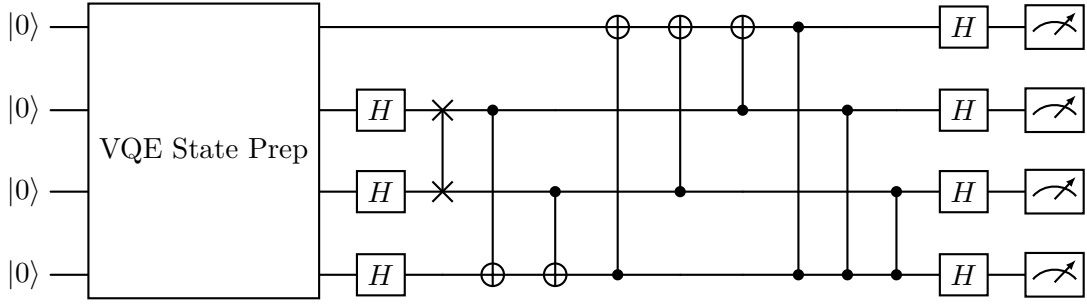
\begin{figure*}
\begin{center}
    \begin{quantikz}
       \ket{0} \ &\gate[4]{\text{VQE State Prep}} &         &           &           &          & \targ{}    & \targ{}   & \targ{}    &\ctrl{3}    &          &           &\gate{H}&\meter{}\\
       \ket{0} \ &                                &\gate{H} & \swap{1}  &  \ctrl{2} &          &            &           & \ctrl{-1}  &            &\ctrl{2}  &           &\gate{H}&\meter{}\\
       \ket{0} \ &                                &\gate{H} & \targX{}   &          & \ctrl{1} &            & \ctrl{-2} &            &            &          & \ctrl{1}  &\gate{H}&\meter{}\\ 
       \ket{0} \ &                                &\gate{H} &           &  \targ{}  & \targ{}  & \ctrl{-3}  &           &            &\control{}  &\control{}&\control{} &\gate{H}&\meter{}\\
    \end{quantikz}
\end{center}
\caption{Example circuit consisting of a state preparation and change into the simultaneous eigenbasis of the Pauli operators $XXXX$, $XXYY$, $XYXY$ and $YXXY$ as presented in Ref.~\cite{gokhale2019minimizing}.
 The state preparation prepares a possible ground state wave function according to some ansatz. The measurement basis is then rotated by a Clifford circuit to a basis where the observables of interest can be measured in parallel.}\label{fig:parallel_meas}
\end{figure*}
As an example, the mapping of the mutually commuting Pauli strings $XXXX$, $XXYY$, $XYXY$, $YXXY$ to the basis $ZIII$, $IZII$, $IIZI$, $IIIZ$ is depicted in Fig.~\ref{fig:parallel_meas}.
With the prior described noise model, with an average Pauli fidelity of $f_{\text{avg}} = 0.996$ on a linear device topology, we find an approximate $\gamma = 12.696$ for regular PEC.
In contrast, our method results in a $\gamma$-factor of 7.335 for pPEC without the XI-reduction as well as $\gamma_{\text{pPEC}_{XI}} = 5.449$ with XI reduction, which is about an order of magnitude improvement in the variance $\gamma^2$.

\subsection{Measurement-based quantum computing}
The application of pPEC within the model of measurement-based quantum computing (MBQC) \cite{briegel2009measurement} is another interesting possibility. 
In MBQC the computation is performed on a large, entangled resource state (in this example a graph state), realized only by Clifford operations 
\begin{equation}
    \ket{G} = \prod_{i,j \in E} {C_Z}_{ij} \ket{+}^{\otimes n}.
\end{equation}

The state is constructed according to a graph $G = (V,E)$, where the vertices correspond to the individual qubits and the edges define which qubits are entangled via $C_Z$ gates.
To perform the computation, the individual qubits are measured in adaptive bases, performed by the transformation $M(\theta) = H R_z(\theta)$ which is equivalent to the implementation of a gate operation on the resource state.
The randomness of a quantum measurement is compensated for by using classically controlled feed-forward operations.

\begin{figure}
    \centering

    \begin{tikzpicture}
        \foreach \x in {0,1,2} {
          \foreach \y in {0,1,2} {
            \node[draw,circle,inner sep=2pt,fill] at (\x,\y) {};
          }
        }    
        \foreach \x in {0,1,2} {
          \foreach \y in {0,1,2} {
            \ifnum\x<2
              \draw (\x,\y) -- (\x+1,\y);
            \fi
            \ifnum\y<2
              \draw (\x,\y) -- (\x,\y+1);
            \fi
          }
        }    

        \foreach \y in {0,1,2} {
          \draw[dotted] (2,\y) -- (2.5,\y);
        }
        
        \foreach \x in {0,1,2} {
          \draw[dotted] (\x,-0.5) -- (\x,0);
        }
        \end{tikzpicture}

\caption{Depiction of the cluster state that can be used as a ressource for universal quantum computations in MBQC.}
\label{fig:cluster_state}
\end{figure}
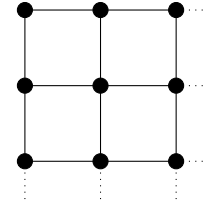

We consider a two-dimensional lattice as shown in Fig.~\ref{fig:cluster_state} as a resource state, which is universal for MBQC \cite{raussendorf2003measurement}.
Such a graph state can be implemented in constant depth, with a minimum depth of $4$ if the device is sufficiently connected.
We again consider a device with a limited, linear connectivity and an average Pauli fidelity of $f_{\text{avg}} = 0.996$.
For the circuit to implement a graph state described by a $4 \times 4$ lattice (transpiled onto the hardware connectivity) we find for regular PEC a $\gamma$-factor of approximately $\gamma_\text{PEC} \approx 4899.6$.
Using pPEC we are able to reduce this factor to approximately $\gamma_{\text{pPEC}_{XI}} \approx 839.1$.
For improved fidelities $f_{\text{avg}} = 0.9996$ and a $7 \times 7$ graph we find $\gamma_\text{PEC} \approx 418.8$ and $\gamma_{\text{pPEC}_{XI}} \approx 324.9$ which is a still considerable improvement of about 20\%.

Our method is  especially well suited for MBQC interesting since the resource state contains all noisy two-qubit operations while only consisting of Clifford gates.
Thus, pPEC including the XI-reduction can be applied to the resource state preparation yielding favorable scaling in comparison to PEC.
Further, pPEC can be applied to correct for the readout errors that occur during the measurement process.
Note, that due to the feed-forward operation, assignment matrix methods can in general not be utilized, rendering a PEC approach more practical~\cite{gupta2024probabilistic}.
Considering the structure of the circuits and the fact that symmetrized measurement errors are defined by an $X$-channel, we can propagate these errors to the start of the circuit. For example an $X$-correction before the measurement can be propagated behind the basis change operation $M(\theta)$ since
\begin{equation}
     M(\theta) X  = R_z(\theta) H X = ZM(\theta), 
\end{equation}
where we utilized that Pauli $Z$ commutes with the $R_z(\theta)$ gate.
Measurement errors can thus be freely propagated throughout the circuit, enabling their incorporation into the pPEC workflow.

\subsection{Further possible applications}
Apart from the circuits discussed, the method can be applied whenever Clifford circuits are present in circuits of practical interest. Potential algorithms include Clifford preconditioning of circuits \cite{sun2024toward} and applications of instantaneous quantum polynomial (IQP) circuits \cite{shepherd2009temporally}, utilizing partial propagation of Pauli errors that commute with the diagonal gates in the IQP circuit. 
While we do not explicitly investigate these approaches in this work, we consider them as promising avenues for further research.

\section{Beyond Clifford Circuits}

\begin{figure*}[hpt!]
\begin{center}
    \begin{quantikz}
        \qw & \gate{R_x(\pm 2h\delta t)}\gategroup[4, steps = 8, style={dotted, cap=round, fill=gray!4.5}, background]{\text{Trotter step, repeated $s$ times}}  & \ctrl{1}&                          &  \ctrl{1} & \qw & \qw & \qw& \qw &\\
        \qw & \gate{R_x(\pm 2h\delta t)} & \targ{} &  \gate{R_z(\mp 2J\delta t)} & \targ{} & \qw & \ctrl{1}&    &  \ctrl{1} &\qw \\
        \qw & \gate{R_x(\pm 2h\delta t)} & \ctrl{1}&   &  \ctrl{1} & \qw &  \targ{} &  \gate{R_z(\mp 2J\delta t)} &   \targ{} & \qw \\
        \qw & \gate{R_x(\pm 2h\delta t)} & \targ{} &  \gate{ R_z(\mp 2J\delta t)} &   \targ{} & \qw & \qw & \qw& \qw &
    \end{quantikz}
\end{center}
\caption{Circuit representation of the time evolution operator of the transverse field Ising model. The circuit shows a single Trotter step $s$ for a model consisting of four qubits which is repeated several times until the circuit is measured. The $\pm$ signs in the circuit indicate that we flip the sign of the rotation angles relative to the Pauli correction that has been sampled.}
\label{fig:IsingCircuit}
\end{figure*}
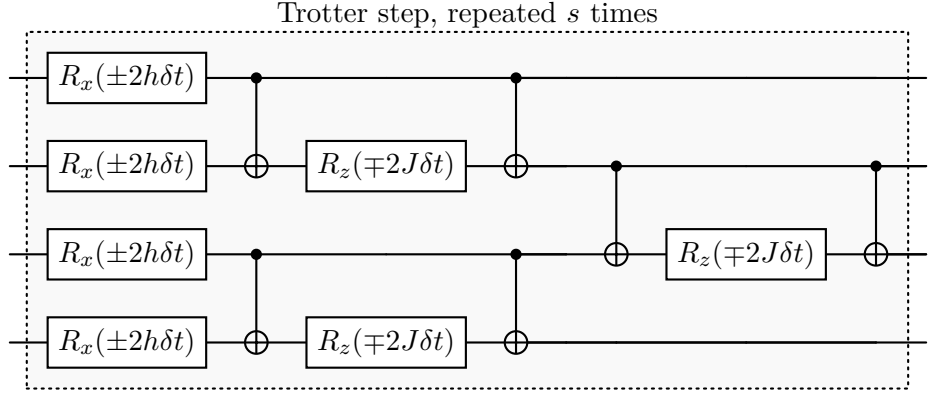

So far, we focused our approach to Clifford circuits. 
This choice is motivated by the fact that Clifford circuits propagate Pauli channels in a well-defined manner without introducing \textit{branching}, allowing for the propagation of channels in deep (Clifford) circuits. 
Branching refers to the phenomenon where the commutation of a Pauli operator with a non-Clifford gate typically results in one Pauli term being mapped to a sum of Pauli terms.
This effect arises from Heisenberg evolution and can be expressed as
\begin{equation}
G^\dagger P G = \sum_{P' \in \{X,Y,Z\}^{\otimes n}}  \text{Tr}(G^\dagger P' G) P'\label{eq:non_pauli}
\end{equation}
where the summation is taken over the Pauli group of dimension $n$.

Generally, pPEC could thus be performed by tracking the evolution of the corrections as sums of Pauli terms.
The problems with this approach are twofold. First, the number of Pauli operators, that need to be tracked, increases exponentially with the number of non-Clifford gates, rendering this method computationally expensive.
Second, single correction operations can be mapped to large Pauli sums with highly non-local operators which cannot be implemented as simple single qubit Pauli gates. A probabilistic implementation of such operators using multi qubit gates might thus introduce significantly more errors into the system.

While both these issues can in principle be mitigated to some extent by either truncating small coefficients in Eq.~\eqref{eq:non_pauli} to keep the Pauli sums small~\cite{angrisani2024classically} or by expressing the channel as a tensor network truncated to bond dimension one to only allow single qubit corrections~\cite{guo2022quantum}, they still pose significant challenges.

An alternative method for handling non-Clifford gates is to restrict the non-Clifford gates to single-qubit Pauli rotation gates $R_{P}(\theta)$, which adhere to the general commutation rule
\begin{equation}
     R_{P'}(\theta) P= P R_{P'}(-1^{\langle P, P' \rangle_\text{SP}} \theta)  
\end{equation}
with $P, P' \in \{I, X,Y,Z\}$ and $\langle P,P' \rangle_\text{SP}$ denoting the symplectic inner product of $P$ and $P'$, which is 0 when the Paulis commute and 1 otherwise.
That is, the commutation will flip the sign of the rotation angle whenever $P$ and $P'$ do not commute.

Although this approach also introduces branching in the context of the rotation angles, a Pauli channel will still be mapped to a Pauli channel, enabling direct sampling of single-qubit Pauli corrections.
Implementing this method requires to keep track of each sign flip for each correction separately and flipping the corresponding signs of the rotation angles before executing the circuit if the correction is sampled.

Since corrections with the same operator but different set of flips do not interfere, this will lead to generally lower reductions.
The main hurdle introduced by keeping track of the signs stems from the fact that the number of possible corrections now scales exponentially in the circuit depth as well as the circuit width.
However, this method might be feasible for low-magic circuits.

\subsection*{Simulation of the transverse-field Ising model}

We conclude the section with considering a specific non-Clifford application, namely the simulation of the transverse-field Ising model using nearest neighbor interactions considered by Ref.~\cite{van2023probabilistic}, which describes magnetic spin-spin interactions with a local transverse field applied to each spin
\begin{equation}
    H_\text{Ising} = -\sum_{j} J_{j,j+1} Z_j Z_{j+1} + \sum_j h_j X_j.
\end{equation}
The time evolution induced by this Hamiltonian can be implemented on a gate-based quantum computer by utilizing the first order Trotter-Suzuki decomposition \cite{trotter1959product, suzuki1976generalized} as
\begin{equation}
    e^{-iHt} \approx (e^{iJ \sum_{j } Z_j Z_{j+1} \delta t}e^{-ih \sum_j X_j \delta t})^s,
\end{equation}
where $\delta t = t/s$  is the total evolution time divided by the number of Trotter steps $s$ and we assume equal coupling among all qubits, i.e., $h_i = h$ and $J_{ij} = J$. 
The implementation of the model on a gate-based quantum computer can then be performed by a set of single site rotation gates $R_X(2h\delta t)$ for the evolution under the transverse-field dynamics and controlled $Z$ rotations, expressed as a rotation gate $R_Z(-2J\delta t)$ between two $C_X$ gates accounting for the spin-spin interactions. We illustrate the circuit for a single Trotter step in Fig.~\ref{fig:IsingCircuit}.
As demonstrated in Ref. \cite{van2022model}, we implement the dynamics of a four qubit Hamiltonian with the parameters $h = 1, J=0.15$ and $\delta t = 1/4$. To estimate the efficiency of our method we measure the magnetization $\mathbf{M} = \sum_i \langle X_i\rangle ,\langle Y_i\rangle, \langle Z_i\rangle / N$, which represents the average polarization along a specified direction.

\begin{figure*}[hpt!]
    \centering
    \makebox[\textwidth][c]{\includegraphics[width=1.2\linewidth]{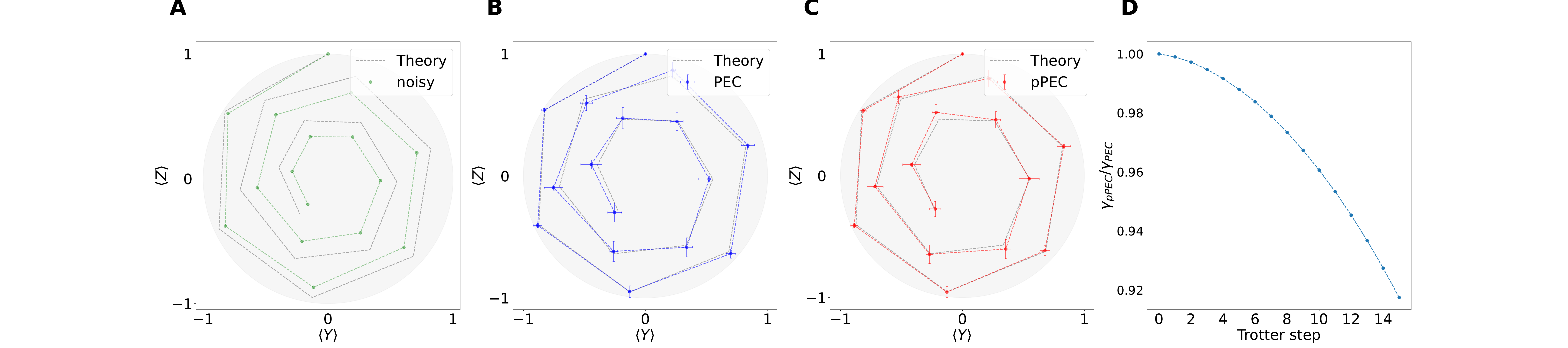}}
    \caption{Dynamics of the average magnetization $\mathbf{M} = \sum_i \langle X_i\rangle ,\langle Y_i\rangle, \langle Z_i\rangle / N$ for a linear chain of 4 qubits for an increasing number of Trotter steps. The results show the noisy simulation (\textbf{A}) the PEC corrected expectation value (\textbf{B}) and the pPEC corrected expectation value (\textbf{C}). In the experiment a total of 200 expectation values per data point has been taken. The error bars show one standard deviation of the estimated data. (\textbf{D}) sampling overhead reduction $\gamma_\text{pPEC} / \gamma_\text{PEC}$ relative to the number of Trotter steps.}
    \label{fig:Magnetization_Plot}
\end{figure*}

We simulate the dynamics over 16 Trotter steps while incorporating gate-level noise. Specifically, we apply a random 2-qubit Pauli error channel with an error probability $p = 0.01$ after each $C_X$ gate. The experiment consists of a noiseless simulation, a noisy simulation, and corrections using PEC and pPEC.
To calculate the pPEC inverse we perform a channel propagation for each inverse channel, tracking the sign of each passed rotation gate for each correction. In the expansion step we truncate the calculation of the channel products by only considering coefficients  $|c_i| > 10^{-6}$ to keep the classical overhead small.

The results, presented in Fig.~\ref{fig:Magnetization_Plot}, illustrate the average magnetization of the four-qubit test circuit in the $y$-$z$ plane. Both PEC and pPEC successfully retrieve the bias-free expectation values within the error bars for all data points. However, unlike the Clifford case, we observe a less significant reduction in the $\gamma$-factor, approximately 10\% for the last Trotter step.
We finally want to highlight that the circuit contains a higher number of non-Clifford gates compared to Clifford gates, making it one of the more challenging circuits for pPEC. 
While the results are far less drastic than the results for Clifford circuits, pPEC is still able to achieve a meaningful reduction.

\section{Discussion and Conclusion}
Quantum error mitigation is a promising avenue to achieve near-term quantum advantage.
Even though tremendous progress in hardware error rates has been achieved, quantum error mitigation is still vital to obtain meaningful results on current hardware.
However, one of the limiting factors for practical error mitigation still lies in the excessive amount of quantum resources required to mitigate large circuits.

In this work we presented a method to, in some cases drastically, reduce the sampling overhead of PEC for Clifford circuits.
PEC in combination with Pauli error propagation is able to retrieve bias free expectation values of noisy circuits. 
The method itself delivers excellent results for current hardware error rates, reducing the sampling overhead by a few orders of magnitude for even moderately deep quantum circuits.
We observed that the effect is far less drastic for reduced error rates, which are expected for future quantum device generations.
However, even for future hardware error rates and quantum circuits consisting of $30$-$50$ qubits a reduction of one to two orders of magnitude can be achieved, making pPEC an attractive choice whenever applicable.

We hope that the promising results of sampling from a fused inverse could spark further insights in the most effective form of canceling quantum noise.
How to optimally estimate and sample from the global inverse is an interesting question for future research.

The clear bottleneck of our algorithm is the classical overhead to calculate the channel products. 
We showed that the expansion can be performed without tremendous overhead by truncation, yielding more limited results.
A further interesting approach to efficiently expand the noise channel could be given by tensor network algorithms, which can approximate a large amount of data with few resources.
By expressing the channels as matrix product operators (MPO) the channel products can be calculated as a contraction of several MPOs~\cite{wood2011tensor, fischer2024dynamical}.
The main problem lies in the fact that, to keep the bond dimension small, approximations need to be made which can lead to a bias in the estimated expectation values.

We presented basic examples of potential applications for our method, where Clifford circuits, despite being efficiently simulatable via the Gottesman-Knill theorem, are of practical interest.
Finally, we tested our method on a non-Clifford example: the simulation of the transverse field Ising model. While the proposed method successfully retrieves the ideal value from noisy data, the reduction in sampling overhead is far lower than that observed for Clifford circuits.
We find the exploration of circuits with low magic or measurement-based quantum computation (MBQC), particularly with respect to qubit recycling, to be especially promising avenues for further research.

\begin{acknowledgements}
This work was in part supported by the research project Zentrum für Angewandtes Quantencomputing (ZAQC), which is funded by the Hessian Ministry for Digital
Strategy and Innovation and the Hessian Ministry of Higher Education, Research and the Arts. We acknowledge further support by the research project Leistungszentrum für innovative Therapeutika (TheraNova), which is funded 
by the Fraunhofer Gesellschaft and the Hessian Ministry for Higher Education, Research and the Arts.
\end{acknowledgements}

\bibliographystyle{unsrtnat}
\bibliography{main}

\onecolumn 
\newpage
\appendix

\section{pPEC via MCMC method}\label{Appendix:MonteCarlo}
Here, we give a description on how to estimate a global inverse noise channel using a Markov chain Monte Carlo (MCMC) simulation.
While in general this approach is not scalable and applicable only to small system sizes it serves as an excellent educational example on how the interference of corrections reduces the sampling overhead.

The key insight to understand how the combination of channels reduces the sampling overhead lies in the interference of corrections.
Considering for example a layer where a specific correction $P_i$ has been sampled, it is possible that the effect of the correction is exactly canceled by a subsequent correction $P_j$, so that the same global correction could be achieved by performing no correction at all.

\begin{figure}[h]
\begin{center}
    \begin{quantikz}
        \qw & \gate{H} & \gate[style={dashed, fill=blue!20}]{I}\gategroup[2,steps=1,style={dashed,rounded
        corners}]{sign $= -1$} & \ctrl{1} & \qw \\
        \qw & \gate{Z} & \gate[style={dashed, fill=blue!20}]{Y} & \targ{}  & \qw
    \end{quantikz}
    $\boldsymbol{\rightarrow}$
    \begin{quantikz}
        \qw & \gate[style={dashed, fill=blue!20}]{I}\gategroup[2,steps=1,style={dashed,rounded
        corners}]{sign $= -1$} & \gate{H} & \ctrl{1} & \qw \\
        \qw & \gate[style={dashed, fill=blue!20}]{Y} & \gate{Z} & \targ{} & \qw
    \end{quantikz} \\[2\baselineskip]

    \begin{quantikz}
        \qw & \gate{H} & \ctrl{1} & \gate[style={dashed, fill=blue!20}]{X}\gategroup[2,steps=1,style={dashed,rounded
        corners}]{sign $= +1$}  & \qw \\
        \qw & \gate{Z} & \targ{} & \gate[style={dashed, fill=blue!20}]{Y} & \qw
    \end{quantikz}
    $\boldsymbol{\rightarrow}$
    \begin{quantikz}
        \qw & \gate[style={dashed, fill=blue!20}]{I}\gategroup[2,steps=1,style={dashed,rounded
        corners}]{sign $= +1$} & \gate{H} & \ctrl{1} & \qw \\
        \qw & \gate[style={dashed, fill=blue!20}]{Y} & \gate{Z} & \targ{} & \qw
    \end{quantikz}
\end{center}
\caption{Example of interfering error paths. The drawn corrections (dashed blue gates) of the individual error paths amount, after the propagation,
to the same correction although at a different sign. Calculating both paths is thus superfluous, since their contribution to the average can be easily accounted for in preprocessing.}
\label{fig:interference}
\end{figure}
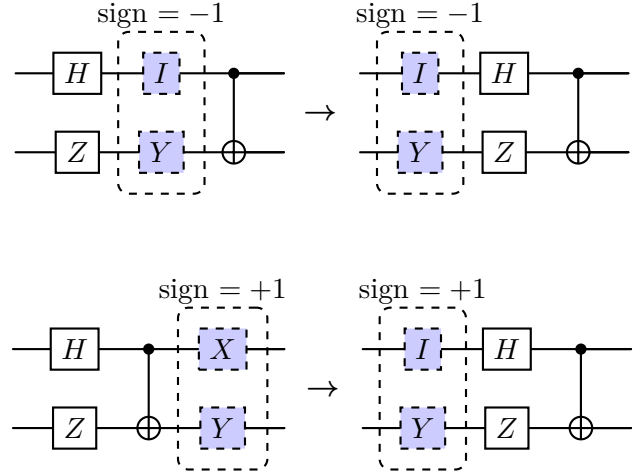

A simple graphical illustration of interfering corrections is shown in Fig.~\ref{fig:interference}.
In this simplified model two corrections are drawn which lead to the same global correction, although at a different global sign.
This pair of paths cancels each other in the average and can be omitted from sampling. 

To quantify the amount of interference we utilize a MCMC algorithm that iterates reversely over each noisy layer in the circuit, drawing a specific Pauli correction $P_l$ and corresponding sign $s_l$ at each step.
Consecutively, the drawn correction is propagated behind the next gate layer $P_l \rightarrow \tilde{P}_l$ using a look-up table (c.f.~Appendix.~\ref{App:Tables}).
At the next gate layer, a new Pauli correction $P_{l-1}$ and sign $s_{l-1}$ are drawn, which get multiplied with the propagated Pauli correction $P_\text{prod} = \tilde{P}_l P_{l-1} $ and the previous sign $s_\text{prod} = s_l s_{l-1} $.
Then the product of the Pauli operators $P_\text{prod}$ is propagated to the next layer, where the same process is repeated until the last noisy layer (or the initialization of the qubits) is reached.

As described in Sec.~\ref{sec:XI-red}, at the initialization or the measurement layer, the corrections $Y$ can be replaced by $X$ and $Z$ by $I$ in each sampled final Pauli correction string via the XI-reduction.

Finally, each time a specific correction $P_\text{final}$ was sampled, we save the number of occurrences and increment or decrement the index of draws depending on $s_\text{final}$.
In this way, some corrections may interfere constructively, while others interfere destructively.
Using this procedure, the reduction in the sampling overhead can be estimated by considering the amount of path interference
\begin{equation}
\label{eq:reduction}
    \frac{\gamma_{\text{ppec}}}{\gamma} = \frac{\text{\#\;paths total} - \text{\#\;interfering paths}}{\text{\#\;paths total}},
\end{equation}
where a \textit{path} refers to one iteration of the MCMC algorithm, in which the corrections have been propagated to the start or end of the circuit and a final global correction has been registered.

It is important to note that, due to the sign problem of the MCMC approach with quasi probability distributions, the number of Monte-Carlo samples needs to be scaled accordingly to address deep circuits. 
\begin{algorithm}[H]
    \caption{Mote-Carlo Simulation to estimate the global inverse}
    \label{pPEC}
    \begin{algorithmic}
    \While{$n < {\rm sample size}$}
        \While{layer index $\neq$ 0}
            \State layer corr, sign $\gets$ sample correction and sign
            \If{layer index = N}
                \State global corr $\gets$ layer corr
                \State global sign $\gets$  sign
                \State Continue
        \EndIf
            \State global corr $\gets$ propagate correction
            \State global corr $\gets$ global corr $\times$ layer corr
            \State global corr sign $\gets$  global sign $\times$ sign
        \State layer index $\gets$ layer index - 1
        \EndWhile
    \State Replace Y with X and Z with I in global correction
    \State add global sign $\times$ global correction to dictionary
    \EndWhile
    \end{algorithmic}
  \end{algorithm}

After the estimation of the global inverse, PEC can be performed by drawing from the estimated distribution obtained from the MCMC simulation.
For a fixed number of circuit instances, a correction and corresponding sign are drawn, and the results are multiplied with the corresponding sign and averaged.
Finally, the results need to be rescaled by the corresponding $\gamma$-factor. 
Due to the path interference, the total $\gamma$ factor now amounts to
\begin{equation}
        \gamma_{\text{ppec}} = \gamma \cdot \frac{\text{\#\;samples} - \text{\#\;interfering samples}}{\text{\#\;samples}},
\end{equation}
showing the same structure as Eq.~\eqref{eq:gamma_ppec}.

Note that due to the exponential scaling of the correction space, this approach requires an exponentially scaling amount of Monte-Carlo samples in order to converge.
To circumvent prohibitive classical costs, the method can be truncated by for example only considering a select number of layers or performing the channel products analytically for a selected set of channels.

\section{Commutation Tables for Clifford Gates}
\label{App:Tables}

In this section we showcase the basic commutation tables for the most used gates of the Clifford group of dimension $n=1$ and $n=2$. 
While it is generally true that each element of the Clifford group can be decomposed into $H$, $S$ and $C_X$ gates we provide commutation tables 
for a larger set of gates, since they are customary used in some decompositions or native gate sets of quantum hardware (such as the basis \{$X$, $S_x$, $R_z$, $C_X$\}).

\begin{table}[ht]
\begin{tabular}{ccc}
\begin{subtable}[t]{0.3\textwidth}
\centering
    \begin{tabular}{|c|c|}
        \hline
        \multicolumn{2}{|c|}{$C_X$-Commutation table}\\
        \hline
        II & II\\ 
        IX & XX\\ 
        IY & XY\\
        IZ & IZ\\ 
        XI & XI\\ 
        XX & IX\\ 
        XY & IY\\
        XZ & XZ\\ 
        YI & YZ\\ 
        YX & ZY\\ 
        YY & ZX\\
        YZ & YI\\ 
        ZI & ZZ\\ 
        ZX & YY\\ 
        ZY & YX\\
        ZZ & ZI\\ 
        \hline
    \end{tabular}
\end{subtable}
\begin{subtable}[t]{0.3\textwidth}
    \begin{tabular}{|c|c|}
        \hline
        \multicolumn{2}{|c|}{$C_Z$-Commutation table}\\
        \hline
        II & II\\ 
        IX & ZX \\ 
        IY & ZY \\
        IZ & IZ \\ 
        XI & XZ\\ 
        XX & YY \\ 
        XY & YX \\
        XZ & XI \\ 
        YI & YZ\\ 
        YX & XY \\ 
        YY & XX \\
        YZ & YI \\ 
        ZI & ZI\\ 
        ZX & IX \\ 
        ZY & IY \\
        ZZ & ZZ \\ 
        \hline
    \end{tabular}    
\end{subtable}
\begin{subtable}[t]{0.3\textwidth}
\centering
\begin{tabular}{|c|c|}
        \hline
        \multicolumn{2}{|c|}{$SWAP$-Commutation table}\\
        \hline
        II & II\\ 
        IX & XI\\ 
        IY & YI\\
        IZ & ZI\\ 
        XI & IX\\ 
        XX & XX\\ 
        XY & YX\\
        XZ & ZX\\ 
        YI & IY\\ 
        YX & XY\\ 
        YY & YY\\
        YZ & ZY\\ 
        ZI & IZ\\ 
        ZX & XZ\\ 
        ZY & YZ\\
        ZZ & ZZ\\ 
        \hline
    \end{tabular}   
\end{subtable}
\end{tabular}
\end{table}

\begin{table}[htp!]
\begin{tabular}{cccc}
\begin{subtable}[t]{0.25\textwidth}
\centering
    \begin{tabular}{|c|c|}
        \hline
        \multicolumn{2}{|c|}{$H$-Commutation table}\\
        \hline
        I & I\\ 
        X & Z \\ 
        Y & Y \\
        Z & X \\ 
        \hline
    \end{tabular}    
\end{subtable}

\begin{subtable}[t]{0.25\textwidth}
\centering
    \begin{tabular}{|c|c|}
        \hline
        \multicolumn{2}{|c|}{$S_X$-Commutation table}\\
        \hline
        I & I\\ 
        X & X\\ 
        Y & Z\\
        Z & Y\\ 
        \hline
    \end{tabular}    
\end{subtable}

\begin{subtable}[t]{0.25\textwidth}
\centering
    \begin{tabular}{|c|c|}
        \hline
        \multicolumn{2}{|c|}{$S_Y$-Commutation table}\\
        \hline
        I & I\\ 
        X & Z \\ 
        Y & Y \\
        Z & X \\ 
        \hline
    \end{tabular}    
\end{subtable}

\begin{subtable}[t]{0.25\textwidth}
\centering
    \begin{tabular}{|c|c|}
        \hline
        \multicolumn{2}{|c|}{$S_Z$-Commutation table}\\ 
            \hline
            I & I\\ 
            X & Y \\ 
            Y & X \\
            Z & Z \\ 
            \hline
        \end{tabular}    
    \end{subtable}
\end{tabular} 
\caption{Commutation tables for the most used gates of the Clifford group. The commutation tables where calculated via the relation $P' = CPC^\dagger$.}
\end{table}

\section{Conjoint implementation of inverse noise channels}\label{Appendix:Proof}
We provide a simple prove that it is always favorable in terms of the sampling overhead to multiply two inverse Pauli channels together before probabilistically implementing them.
We start the proof by considering a general inverse Pauli noise channel
\begin{equation}
    \Lambda^{-1}(\rho) = \sum_{i=1}^{4^n} a_i P_i \rho P_i^\dagger = \sum_{i=1}^{4^n} a_i \mathcal{P}_i \quad a_i \in \mathds{R}
\end{equation}
where $\mathcal{P}_i (\rho) \overset{\scriptscriptstyle\wedge}{=} P_i \rho P_i^\dagger$ denotes the Kraus channel representation of a Pauli $P_i$.
Considering that the sampling overhead $\gamma$ is calculated as
\begin{equation}
    \gamma(\Lambda^{-1})  = \gamma\left( \sum_{i=1}^{4^n} a_i \mathcal{P}_i \right) = \sum_{i=1}^{4^n} |a_i|,
\end{equation}
it is easy to verify that the sampling overhead of implementing two inverse noise channels is given by
\begin{equation}
    \gamma(\Lambda^{-1}_1) \gamma(\Lambda^{-1}_2) = \sum_{i, j=1}^{4^n} |a_i| |b_j|.
\end{equation}

On the other hand, first multiplying both inverse channels and then calculating the sampling overhead of the fused channel, yields
\begin{equation}
\label{eq:proof}
\begin{split}
    \gamma(\Lambda^{-1}_1 \Lambda^{-1}_2) &= \gamma \left(\sum_{i=1}^{4^n} a_i \mathcal{P}_i \sum_{j=1}^{4^n} b_j \mathcal{P}_j \right) \\
                                          &= \gamma\left(\sum_{k=1}^{4^n} \left(\sum_{i,j}^{4^n} a_i b_j \xi_{ijk} \right) \mathcal{P}_k\right) \quad \text{with} \quad
                                           \xi_{ijk} = \begin{cases} 1 & \text{if $\mathcal{P}_i \mathcal{P}_j = \mathcal{P}_k$ }\\
                                            0 & \text{otherwise}
                                            \end{cases}\\
                                          &= \sum_k^{4^n} \left(|\sum_{i,j}^{4^n} a_i b_j \xi_{ijk} |\right)\\
                                          &\leq \sum_k^{4^n} \left(\sum_{i,j}^{4^n}|a_i||b_j| |\xi_{ijk}|\right)\\
                                          &= \sum_{i, j=1}^{4^n} |a_i| |b_j| = \gamma(\Lambda^{-1}_1) \gamma(\Lambda^{-1}_2),
\end{split}
\end{equation}
where we made use of the triangle inequality as well as the fact that the total number of terms $a_i b_j$ needs to be conserved by the summation over $k$, i.e., $\xi_{ijk}$ contains only one non-zero element for each product $\mathcal{P}_i\mathcal{P}_j$, $\sum_k |\xi_{ijk}| = 1$ $\forall i,j$.

Thus, the cost of implementing two inverse channels separately is always greater or equal than the conjoint implementation.
Note that in contrast to taking the product of two Pauli observables, the phase of the product of two Pauli channels can be omitted since $(e^{i \phi} P) \rho (e^{i \phi} P)^{\dagger} =  P \rho P^{\dagger}$.

\section{Symmetrization of readout errors}\label{appendix:readout}
We consider a readout error, which is described by a tensor product noise model
\begin{equation}   
        A = \bigotimes_{i=1}^n A^i = \bigotimes_{i=1}^n \begin{bmatrix}
            1-\epsilon_i & \eta_i \\
            \epsilon_i & 1-\eta_i 
            \end{bmatrix},
\end{equation}
such that the errors on each qubit occur independently.
The measurement is described by a set of projectors $\pi_m = \ket{m}\bra{m}$, which project onto the computational basis.
The noisy projection operators are then given by 
\begin{equation}
\tilde{\pi}_m = \sum_n A_{mn} \ket{n}\bra{n}
\end{equation}
where $A_{mn} = \bra{m}A\ket{n}$ are the matrix elements of the error matrix.
For the considered noise model the matrix $A$ is sparse with only few non-vanishing matrix elements.
The randomized insertion of $X$ gates and subsequent correction by post-processing leads to a twirled readout error~\cite{van2022model}, which is given by 
\begin{equation}
    A_\text{twirl} = \frac{1}{2^n} \sum_{s \in \{0,1\}^n} X^{\otimes s} A X^{\otimes s, \dagger},
\end{equation}
where the $X^{\otimes s}$ operator corresponds to bit flips of the qubits indexed by $s$.

Since the noise model does not contain contributions of cross talk it suffices to consider each single qubit error $A^i$ twirled under a single $X$ separately
\begin{equation}
    \begin{split}
        A_\text{twirl} ^i &= \frac{1}{2}( A^i + X A^i X^\dagger) \\
                          &= \frac{1}{2}\left( \begin{bmatrix} 1-\epsilon_i & \eta_i \\
                                               \epsilon_i & 1-\eta_i \end{bmatrix} 
                                            +  \begin{bmatrix} 0 & 1 \\
                                               1 & 0 \end{bmatrix} 
                                               \begin{bmatrix} 1-\epsilon_i & \eta_i \\
                                               \epsilon_i & 1-\eta_i \end{bmatrix} 
                                               \begin{bmatrix} 0 & 1 \\
                                               1 & 0 \end{bmatrix}                         
                                            \right)\\
                        &= \frac{1}{2}\left( \begin{bmatrix} 1-\epsilon_i & \eta_i \\
                                            \epsilon_i & 1-\eta_i \end{bmatrix} 
                                            +  
                                            \begin{bmatrix} 1-\eta_i & \epsilon_i \\
                                            \eta_i & 1-\epsilon_i \end{bmatrix}                        
                                            \right)\\
                        &=     \begin{bmatrix} 1-p_x & p_x \\
                                                p_x & 1-p_x \end{bmatrix}                        
                                            \\  
                        &=   (1-p_x)  \begin{bmatrix} 1 & 0 \\
                             0 & 1 \end{bmatrix}
                             + p_x \begin{bmatrix} 0 & 1 \\
                                1 & 0 \end{bmatrix}\\                  
    \end{split}
\end{equation}
with $p_x = \frac{\epsilon_i + \eta_i}{2}$.
Hence the randomized insertion of $X$ gates transforms a general single qubit readout error channel into a Pauli $X$ error channel.

We would like to note, that the method straight-forwardly generalizes to either the full readout error matrix or a model considering crosstalk of a subset of qubits.
The integration of larger matrices comes at the cost of estimating each correlated measurement, but might generally reduce or even fully eliminate any bias.

\section{pPEC for SPL noise models}\label{Appendix:SPL}
We now consider pPEC applied to sparse Pauli-Lindbladian (SPL) noise models.
These models can be efficiently stored in a product representation of individual model Paulis which only model interactions between physically connected qubits.
We depart from the product representation of an inverse Pauli channel as given in Ref.~\cite{van2023probabilistic}:
\begin{equation}
\label{eq:SPL_noise_model}
    \Lambda^{-1} (\rho) = \prod_{k \in \mathcal{K}}(w_k \mathcal{I}(\cdot) - (1-w_k) \mathcal{P}_k(\cdot)) \rho.
\end{equation}

Akin to the dense representation, the quasi probability implementation of the inverse channel is attached to a cost factor $\gamma$.
For the SPL model the $\gamma$-factor is given by 
\begin{equation}
    \gamma = \prod_{k \in \mathcal{K}}(2 w_k - 1)^{-1},
\end{equation}
which we aim to reduce using pPEC.
Applying the pPEC method, each of the individually commuting Pauli channels are propagated to the start of the circuit by taking the conjugation with the circuit operation up to the select inverse channel,
\begin{equation}
\begin{split}
    \tilde{\Lambda}^{-1} (\rho) &= \prod_{k \in \mathcal{K}}(w_k \mathcal{C}(\mathcal{I})(\cdot) - (1-w_k) \mathcal{C}(\mathcal{P}_k)(\cdot)) \rho\\
                           &= \prod_{k \in \mathcal{K}}(w_k \mathcal{I}(\cdot) - (1-w_k) \tilde{\mathcal{P}_k}(\cdot)) \rho,
\end{split}
\end{equation}
where $\mathcal{C}$ denotes the Clifford circuit operation up to the respective inverse error channel and the operator $\tilde{\mathcal{P}_k} = C P_k C^\dagger \cdot C^\dagger P_k^\dagger C$ denotes the conjugated Pauli channel.

Applying this procedure to each correction layer, the global inverse is given by 
\begin{equation}
\label{eq:prod}
    \Lambda_{\text{global}}^{-1} = \prod_l \tilde{\Lambda}_l^{-1} = \prod_{k \in L \cdot \mathcal{K}}(w_k \mathcal{I}(\cdot) - (1-w_k) \tilde{\mathcal{P}_k}(\cdot)) \rho
\end{equation}
where the index $k$ now runs over all model Paulis for the total number of layers $L$, yielding a total of $L \cdot |\mathcal{K}|$ terms.
Again, as in the MCMC approach, a further reduction can be achieved by utilizing the XI-reduction. 
In this model it suffices to replace the operators $Y$ by $X$ and $Z$ by $I$ in each of the individual Pauli channels in the product.
As an immediate improvement all channels, which contain only $Z$ and $I$ Pauli operators can be directly omitted from the inverse noise channel and do not contribute to the sampling overhead, since
\begin{equation}
    (w_k \mathcal{I} - (1-w_k)\mathcal{I}) = (2 w_k - 1) \mathcal{I}
\end{equation}
directly cancels with the corresponding term $(2 w_k - 1)^{-1}$ in the $\gamma$ factor.
We denote this as \textit{passive reduction}.

The global inverse noise channel will in general contain multiple Paulis that have been mapped to the same operator.
The amount of terms in the product can then be directly reduced by multiplying terms with equal Paulis together
\begin{equation}
    \begin{split}
        (w_1 \mathcal{I} - (1-w_1) \mathcal{P}) \cdot (w_2 \mathcal{I} - (1-w_2) \mathcal{P}) &= (w_1 w_2 + (1-w_1)(1-w_2)) \mathcal{I} - (w_1(1-w_2)) + (w_2(1-w_1)) \mathcal{P}\\
                                                &= w_3 \mathcal{I} - (1-w_3) \mathcal{P}.
    \end{split},
\end{equation}
with $w_3=(w_1 w_2 + (1-w_1)(1-w_2))$.
Since
\begin{equation}
    (2w_1 - 1)^{-1}(2w_2 - 1)^{-1} = (2 w_3 - 1)^{-1}, 
\end{equation}
this multiplication does not reduce the sampling overhead $\gamma$ but can drastically reduce the number of total terms in the product.

As explained in Ref.~\cite{van2023probabilistic} a further decrease in the sampling overhead can be achieved by explicitly expanding the product of Eq.~\eqref{eq:prod}
\begin{equation}
    \Lambda^{-1} (\rho) = \prod_{k \in \mathcal{K}}(w_k \mathcal{I}(\cdot) - (1-w_k) \mathcal{P}_k(\cdot)) \rho \rightarrow  \sum_i c_i \mathcal{P}_i(\rho).
\end{equation}
Due to the multiplication of equal Pauli channels as well as the larger number of terms in this product, this expansion will yield far more drastic reductions than the expansion of an individual inverse channel given by Eq.~\eqref{eq:SPL_noise_model}. 

In the worst case, the expansion will result in an exponential growth in the number of required parameters to characterize the noise channel.
This leads to a general trade-off between required classical and quantum resources.
However, the growth in the number of required coefficients can be reduced by utilizing specific term orderings before the expansion.

\section{Efficient expansion of SPL noise models}\label{EfficientExpansion}
As stated in Ref.~\cite{van2023probabilistic} the product structure of the noise model in SPL form can be explicitly expanded to reduce the sampling overhead $\gamma$, although at a computational cost in compute time as well as memory.

The number of required parameters and multiplications for the model scales quasi-exponentially, meaning that each term, that is absorbed into the sum, can in theory double the number of required coefficients.
To evade this exponential increase, the expansion can be truncated cutting the multiplication off at a selected index
\begin{equation}
    \Lambda^{-1}(\rho) = \prod_{k' \in \mathcal{K'}}(w_{k'} \mathcal{I}(\cdot) - (1-w_{k'}) \mathcal{P}_{k'}(\cdot)) \left(\sum_i c_i \mathcal{P}_i(\cdot) \right) \rho.
\end{equation}
The reduction in the sampling overhead can then be estimated by calculating the $\gamma$-factor of the expanded inverse channel
\begin{equation}
    \gamma_{\text{reduction}} = \sum_i |c_i| \leq 1,
\end{equation}
which is less or equal to 1 due to interference.
While this approach is generally computationally more bearable, a direct expansion of the product, without exploiting structure, is still taxing and often inefficient. 
This inefficiency can be explained by considering a single expansion step
\begin{equation}
    (w_i \mathcal{I} - (1-w_i) \mathcal{P}_i) \left(\sum_k c_k \mathcal{P}_k \right) = \left(w_i \sum_k c_k \mathcal{P}_k \mathcal{I} - (1-w_i) \sum_k c_k \mathcal{P}_k \mathcal{P}_i \right),
\end{equation}
which leads to two separate sums, doubling the number of required coefficients.
The amount of coefficients will however only double if no terms in the two sums are equal, that is no terms interfere.
Luckily, this case is not of interest for pPEC, since the reduction in sampling overhead exactly stems from this interference.
\begin{figure}
    \begin{subfigure}[h]{0.49\linewidth}
    \includegraphics[width=\linewidth]{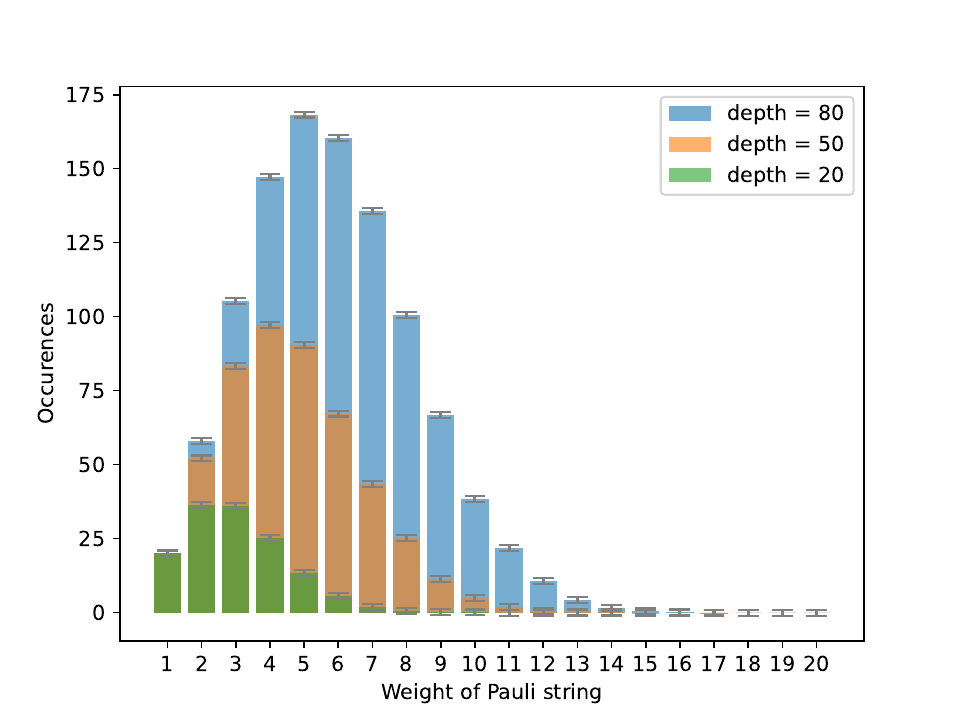}
    \caption{}
    \end{subfigure}
    \hfill
    \begin{subfigure}[h]{0.49\linewidth}
    \includegraphics[width=\linewidth]{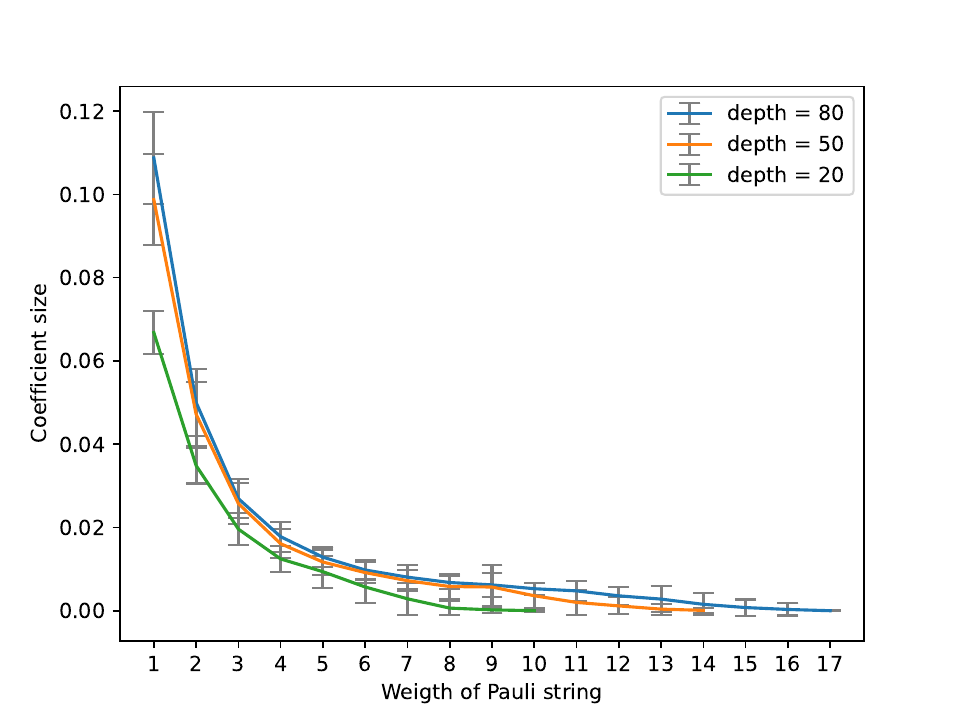}
    \caption{}
    \end{subfigure}%
    \caption{Absolute value of coefficients and weights of the model Paulis for different circuit depths. The data is averaged over several instances of random Clifford circuits of $20$ qubits. 
    (a) Amount of Pauli strings relative to the number of non identity terms. For deeper circuits more Paulis with a higher weight are present.
    (b) Absolute value of the coefficient $|(1-w_k)|$ of the different Pauli strings after propagation relative to the amount of non identity terms in the Pauli string.
}
\label{fig:Pauli_distrib}
\end{figure}

One approach to maximize the amount of interference is the utilization of the subgroup structure of the Pauli operators.
As a basic example, consider the expansion of the three single qubit Pauli channels 
\begin{equation}
    (w_1 \mathcal{I} - (1-w_1) \mathcal{X})(w_2 \mathcal{I} - (1-w_2) \mathcal{Y})(w_3 \mathcal{I} - (1-w_3) \mathcal{Z}) = (1-p_1 - p_2 - p_3 ) \mathcal{I} + p_1 \mathcal{X} + p_2 \mathcal{Y} + p_3 \mathcal{Z}
\end{equation}    
which can be expanded without any increase in the number of coefficients.
The cursive operators indicate that the operators act as channels in Kraus representation i.e. $\mathcal{X} = X(\cdot)X^\dagger$.
The same holds trivially for any larger subgroup, as for example the 15 parameter group of $n=2$, the 63 parameter group of $n=3$ as well as any higher dimensional subgroup.

Further, the propagation of Paulis through deep circuits will lead to high weight Paulis as demonstrated in Fig.~\ref{fig:Pauli_distrib} (a).
On the other hand, the size of the coefficients $w_k$ is far smaller than those of low weight operators (Fig.~\ref{fig:Pauli_distrib} (b)) since it is highly unlikely that two corrections are mapped to the same high weight operator.
Since the general reduction of the expansion is dependent on the magnitude of the channel coefficients, the absorption of low weight Paulis is more favorable.

Based on these heuristics we define a more efficient expansion of the product as follows.
We start by ordering the terms in lexicographical ordering~\cite{funcke2022measurement}
\begin{equation}
    \begin{split}
            III &\preceq IIX \preceq IIY \preceq IIZ\\
    \preceq IXI &\preceq IXX \preceq IXY \preceq IXZ\\
    \preceq IYI &\preceq IYX \preceq IYY \preceq IYZ \preceq ...\\
    \end{split}
\end{equation}
and expanding the product of the first two terms. 
In each consecutive step we absorb a term only if its support is also in the continuously growing expanded inverse noise channel.
If no Pauli with the same support is found, the next Pauli in the ordering is chosen.

Generally, after expanding the first truncated sum, the process can be iteratively applied to the remainder of the product, yielding several partially expanded sums 
\begin{equation}
    \Lambda^{-1} (\rho) = \prod_{k \in \mathcal{K}}(w_k \mathcal{I}(\cdot) - (1-w_k) \mathcal{P}_k(\cdot)) \rho \rightarrow \left(\sum_{i_1} c_{i_1} \mathcal{P}_{i_1}(\cdot) \right) \left(\sum_{i_2} c_{i_2} \mathcal{P}_{i_2}(\cdot) \right) ... \left(\sum_{i_N} c_{i_N} \mathcal{P}_{i_N}(\cdot) \right) \rho.
\end{equation}
Using this process the inverse channel can be expanded into multiple smaller sums, each with only a limited number of terms and total sampling overhead
\begin{equation}
    \gamma_{\text{pPEC}} = \gamma \prod_{j=1}^N \sum_i |c_{i_j}|.
\end{equation}
While we do find that the lexicographical ordering leads to significant reductions it is possible that other orderings might lead to even more interference.

\end{document}